\documentclass[twocolumn,showpacs,aps,prd,xspace,nofootinbib]{revtex4}

\usepackage{graphicx}
\usepackage{dcolumn}
\usepackage{amsmath}
\usepackage{amssymb}
\usepackage{epsfig}

\input babarsym

\newcommand{\mmiss}{\ensuremath{m_\mathrm{miss}^2}\xspace}
\newcommand{\ds}{\ensuremath{D^{(*)}}\xspace}
\newcommand{\eextra}{\ensuremath{E_\mathrm{extra}}\xspace}
\newcommand{\btag}{\ensuremath{B_\mathrm{tag}}\xspace}
\newcommand{\pstarl}{\ensuremath{|{\bf p}^*_\ell|}\xspace}
\newcommand{\gevccnosp}{\ensuremath{{\mathrm{Ge\kern -0.1em V\!/}c^2}}\xspace}
\newcommand{\nll}{\ensuremath{-\log\mathcal L}\xspace}

\begin{document}


\begin{flushleft}
\end{flushleft}

\title{
{\large \bf
\boldmath Measurement of the Semileptonic Decays $B\to D\taum\nutb$ and $B\to\Dstar\taum\nutb$}
}

%
\author{B.~Aubert}
\author{M.~Bona}
\author{Y.~Karyotakis}
\author{J.~P.~Lees}
\author{V.~Poireau}
\author{E.~Prencipe}
\author{X.~Prudent}
\author{V.~Tisserand}
\affiliation{Laboratoire de Physique des Particules, IN2P3/CNRS et Universit\'e de Savoie, F-74941 Annecy-Le-Vieux, France }
\author{J.~Garra~Tico}
\author{E.~Grauges}
\affiliation{Universitat de Barcelona, Facultat de Fisica, Departament ECM, E-08028 Barcelona, Spain }
% ...
\collaboration{The \babar\ Collaboration}
\noaffiliation

\date{\today}

\begin{abstract}
We present measurements of the semileptonic decays 
$\Bm\to\Dz\taum\nutb$, $\Bm\to\Dstarz\taum\nutb$,
$\Bzb\to\Dp\taum\nutb$, and $\Bzb\to\Dstarp\taum\nutb$,
which are sensitive to non--Standard Model amplitudes in
certain scenarios.
The data sample consists of $232\times 10^6\ \FourS\to\BB$ decays
collected with the \babar\ detector at the \pep2 \epem collider.
We select events with a $D$ or \Dstar meson and a light lepton ($\ell=e$ or $\mu$)
recoiling against a fully reconstructed $B$ meson.
We perform a fit to the joint distribution of
lepton momentum and missing mass squared to distinguish
signal $B\to\ds\taum\nutb\ (\taum\to\ellm\nulb\nut)$
events from the backgrounds, predominantly $B\to\ds\ellm\nulb$.
We measure the branching-fraction ratios
$R(D)\equiv\BR(B\to D\taum\nutb)/\BR(B\to D\ellm\nulb)$ and
$R(\Dstar)\equiv\BR(B\to\Dstar\taum\nutb)/\BR(B\to\Dstar\ellm\nulb)$ and,
from a combined fit to \Bm and \Bzb channels,
obtain the results $R(D)=(41.6\pm 11.7\pm 5.2)\%$ and
$R(\Dstar)=(29.7\pm 5.6\pm 1.8)\%$, where the uncertainties
are statistical and systematic. Normalizing to measured
$\Bm\to D^{(*)0}\ellm\nulb$ branching fractions, we obtain
$\BR(B\to D\taum\nutb)=(0.86\pm 0.24\pm 0.11\pm 0.06)\%$ and
$\BR(B\to\Dstar\taum\nutb)=(1.62\pm 0.31\pm 0.10\pm 0.05)\%$,
where the additional third uncertainty is from the normalization mode.
We also present, for the first time, distributions
of the lepton momentum, \pstarl, and the squared momentum transfer,
$q^2$.
\end{abstract}

\pacs{12.15.Hh, 13.20.-v, 13.20.He, 14.40.Nd, 14.80.Cp}

\maketitle

\section{Introduction}
Semileptonic decays of $B$ mesons to the $\tau$ 
lepton---the heaviest of the three charged leptons---provide 
a new source of 
information on Standard Model (SM) 
processes~\cite{KornerSchuler,Falk_etal,HuangAndKim}, 
as well as a new window on physics beyond 
the SM~\cite{GrzadkowskiAndHou,Tanaka,KiersAndSoni,Itoh,ChenAndGeng,KamenikMescia}.
In the SM, semileptonic decays occur at tree level and are mediated by the $W$ boson, but
the large mass of the $\tau$ lepton provides sensitivity to 
additional amplitudes, such as those mediated by a charged Higgs 
boson. Experimentally, $b\to c\taum\nutb$ decays~\footnote{Charge-conjugate
modes are implied throughout.} are 
challenging to study because the
final state contains not just one, but two or three neutrinos as a
result of the $\tau$ decay.

Theoretical predictions for semileptonic decays to exclusive final states 
require knowledge of the form factors, which parametrize the hadronic current
as functions of $q^2=[p_B-p_{\ds}]^2.$ For light leptons $\ell\equiv e,\ \mu$,\footnote{Throughout
this article, we use the symbol $\ell$ to refer only to the light charged
leptons $e$ and $\mu$.} there is effectively
one form factor for $B\to D\ellm\nulb$, while there are 
three for $B\to\Dstar\ellm\nulb$. If a $\tau$ lepton is produced instead,
one additional form factor enters in each mode. The form factors for $B\to\ds\ellm\nulb$ 
decays~\footnote{The symbol \ds refers either to a $D$ or a \Dstar meson.}
involving the light leptons have been measured~\cite{cleoFF,babarFF,babarFF2},
providing direct information on four of the six form factors.
Heavy quark symmetry (HQS) relations~\cite{IsgurWise} allow one to express
the two additional form factors for $B\to\ds\taum\nutb$ in terms of the
form factors measurable from decays with the light leptons. With sufficient data, one could
probe the additional form factors and test the HQS relations.

Branching fractions for semileptonic $B$ decays to $\tau$ leptons are predicted to be
smaller than those to light leptons.
Calculations based on the SM predict
${\cal B}(\Bzb\to\Dp\taum\nutb)=(0.69\pm0.04)\%$ and
${\cal B}(\Bzb\to\Dstarp\taum\nutb)=(1.41\pm0.07)\%$~\cite{ChenAndGeng},
which account for most of the predicted inclusive rate
${\cal B}(B\to X_c\taum\nutb)=(2.30\pm0.25)\%$~\cite{Falk_etal}
(here, $X_c$ represents all
hadronic final states from the $b\to c$ transition).
In multi-Higgs doublet
models~\cite{GrzadkowskiAndHou,Tanaka,KiersAndSoni,Itoh,ChenAndGeng}, substantial
departures, either positive or negative, from the SM decay rate could occur for
${\cal B}(B\to D\taum\nutb)$, while smaller departures are expected
for ${\cal B}(B\to\Dstar\taum\nutb)$.
Thus, measurements of ${\cal B}(B\to D\taum\nutb)$ are
more sensitive to non-SM contributions than either
${\cal B}(B\to\Dstar\taum\nutb)$ or the inclusive rate.
In addition to the branching fractions, several other
observables are sensitive to possible non-SM contributions,
including $q^2$ distributions and \Dstar and $\tau$
polarization~\cite{GrzadkowskiAndHou,Tanaka,KiersAndSoni,ChenAndGeng,NTW}.

The first measurements of semileptonic $b$-hadron decays to $\tau$ leptons
were performed by 
the LEP experiments~\cite{LEP} operating at the \Z resonance,
yielding an average~\cite{PDG} inclusive branching fraction
${\cal B}(b_{\rm had}\to X\taum\nutb)=(2.48\pm 0.26)\%$, where $b_{\rm had}$ 
represents the mixture of
$b$-hadrons produced in $\Z\to\bbbar$ decays. The Belle experiment has
reported ${\cal B}(\Bzb\to\Dstarp\taum\nutb)=(2.02^{+0.40}_{-0.37}\pm0.37)\%$ 
\cite{Belle}.

The \babar\ Collaboration has presented a measurement of the branching
fractions for $B\to D\taum\nutb$ and $B\to\Dstar\taum\nutb$ for both
charged and neutral $B$ mesons~\cite{prl}. In this article, we
describe the analysis in greater detail, with particular emphasis on
several novel features of
the event selection and fit technique.
We also present distributions of two important kinematic variables,
the lepton momentum, \pstarl, and the squared momentum transfer, $q^2$.

\subsection{Analysis overview and strategy}
We determine the branching fractions of
four exclusive decay modes: $\Bm\to\Dz\taum\nutb$,
$\Bm\to\Dstarz\taum\nutb$, $\Bzb\to\Dp\taum\nutb$, and $\Bzb\to\Dstarp\taum\nutb$,
each of which is measured as a branching-fraction ratio $R$ relative to the corresponding $e$ and $\mu$
modes. To reconstruct the $\tau$, we use the decays
$\taum\to\en\nueb\nut$ and $\taum\to\mun\numb\nut$, which are experimentally the most
accessible. The main challenge of the
measurement is to distinguish $B\to\ds\taum\nutb$ decays, which have
three neutrinos, from $B\to\ds\ellm\nulb$ decays, which have
the same observable final-state particles but only one neutrino.

The analysis strategy is to reconstruct the decays of both $B$ mesons in the $\FourS\to\BB$ 
event, providing powerful constraints on unobserved particles. One $B$ 
meson, denoted \btag, is fully reconstructed in a purely hadronic decay chain. The remaining
charged particles and photons are required to be consistent with the products of a $b\to c$
semileptonic $B$ decay: the daughter charm meson (either a $D$ or \Dstar)
and a lepton ($e$ or $\mu$). The lepton may be either primary or from
$\taum\to\ellm\nulb\nut$.
To distinguish signal events from the normalization modes $B\to\ds\ellm\nulb$, we calculate the missing four-momentum, 

\begin{equation}
p_\mathrm{miss}=p_{\epem}-p_{\rm tag}-p_{\ds}-p_{\ell}
\end{equation}

\noindent of any particles recoiling against the 
observed $\btag+\ds\ell$ system. A large peak at zero in $\mmiss= p_\mathrm{miss}^2$ corresponds 
to semileptonic decays with one neutrino, whereas signal events produce a broad tail out to 
$\mmiss\sim 8\ (\gevccnosp)^2$.

To separate signal and background events, we 
perform a fit (described in Section~\ref{sec:fit})
to the joint distribution of \mmiss and the lepton 
momentum (\pstarl) in the rest frame of the $B$ meson.
In signal events, the observed lepton is the daughter of the $\tau$ 
and typically has a soft spectrum; for most background events, this lepton 
typically has higher momentum. The fit is performed simultaneously in
eight channels, with a set of constraints relating the event yields
between the channels. The fit is designed to maximize the sensitivity
to the $B\to D\taum\nutb$ signals by using events in the $\Dstar\ellm$ channels
to constrain the dominant backgrounds, $B\to\Dstar\taum\nutb$
feed-down, in which the final-state \Dstar meson is not completely reconstructed. Similarly, we use a set of $D^{**}$ control samples
to constrain the feed-down background to both the $D\taum\nutb$ and
$\Dstar\taum\nutb$ signals.\footnote{Throught this paper,
we use the symbol $D^{**}$ to
represent all charm resonances heavier than the $\Dstar(2010)$, as well as
non-resonant $\ds n\pi$ systems with $n\geq 1$.}

We perform a relative measurement, extracting both signal $B\to\ds\taum\nutb$
and normalization $B\to\ds\ellm\nulb$ yields from the fit to obtain the 
four branching-fraction ratios $R(\Dz)$, $R(\Dp)$, $R(\Dstarz)$, and $R(\Dstarp)$, where,
for example, $R(\Dstarz)\equiv{\cal B}(\Bm\to\Dstarz\taum\nutb)/{\cal B}(\Bm\to\Dstarz\ellm\nulb)$. 
In the ratio, many systematic uncertainties cancel, either partially or
completely.
These ratios are normalized such that $\ell$ represents only one of $e$ or $\mu$;
however, both light lepton species are included in the measurement.
We multiply these branching-fraction ratios by previous measurements of
$\BR(B\to\ds\ellm\nulb)$ to derive absolute branching fractions.

\section{The \babar\ Detector and Data Sets}
We analyze data collected with the \babar\ detector
at the \pep2 \epem storage rings at the Stanford Linear Accelerator Center.
\pep2 is an asymmetric-energy $B$ factory, colliding
$9.0\ \gev\ e^-$ with $3.1\ \gev\ e^+$ at a center-of-mass
energy of $10.58\ \gev$, corresponding to the \FourS
resonance.
The data sample used
consists of $208.9\ \invfb$ of integrated luminosity recorded on the $\FourS$
resonance between 1999 and 2004, yielding $232\times 10^6\ \FourS\to\BB$ decays.
This data sample can be divided into two major periods: Runs 1--3, 
comprising $109.0\ \invfb$ taken
from 1999 to June 2003, and Run 4, comprising $99.9\ \invfb$ taken
from September 2003 to July 2004. The
accelerator background conditions were significantly different between
Runs 1--3 and Run 4, which could affect missing-energy
analyses such as this one; for this reason, the two running periods
have been independently validated, and the fraction of signal-like
events found in the Run 4 sample is used as a crosscheck of the
results, as described in Section~\ref{results}.

The \babar\ detector is a large, general-purpose magnetic
spectrometer and is described in detail elsewhere~\cite{BABARNIM}.
Charged particle trajectories are measured in a tracking
system consisting of a five-layer double-sided silicon strip
detector and a 40-layer drift chamber, both of which operate
in the $1.5\ \mbox{T}$ magnetic field of a superconducting solenoid. A detector
of internally reflected Cherenkov light (DIRC) is used to
measure charged particle velocity for particle
identification (PID).
An electromagnetic calorimeter (EMC), consisting of 6580 CsI(Tl) crystals,
is used to reconstruct photons and in electron identification.
The steel flux return of the solenoid is segmented and instrumented
with resistive plate chambers (IFR) for muon and neutral hadron identification.

All detector systems contribute to charged particle identification.
Ionization energy losses in the tracking
systems and the Cherenkov light signature in the DIRC
are used for all charged particle types. Electrons are also
identified on the basis of shower shape in the EMC and the ratio
of energy deposited in the EMC to the track momentum.
Muon identification is based on a minimum-ionization energy deposit
in the EMC and on the measured interaction length in the IFR.

This analysis relies on measurement of the missing momentum
carried off by multiple neutrinos, and the large solid angle coverage (hermeticity) of the
detector is therefore crucial. The tracking system, calorimeter, and
IFR cover the full azimuthal range and the polar angle range
from approximately $0.3<\theta<2.7\rad$ in the laboratory frame,
corresponding to a \FourS center-of-mass
coverage of approximately 90\% (the direction $\theta=0$
corresponds to the direction of the high-energy beam, and therefore
to the \FourS boost). The DIRC fiducial volume is
slightly smaller, corresponding
to a center-of-mass frame coverage of about 84\%.

Within the active detector volume, the efficiency for reconstructing
charged tracks and photons is very high, typically greater than 95\% over
most of the momentum range. At low momenta, however, the reconstruction
efficiency drops off, leading to an increased contribution from feed-down processes to which special
attention is paid throughout this analysis. Feed-down occurs when
the photon from $\Dstar\to D\gamma$ or the \piz from
$\Dstar\to D\piz$ is not reconstructed (in the case of the \piz,
either one or both of the photons from $\piz\to\gamma\gamma$ may
be missed). Care must therefore be taken to avoid confusing
\Dstar feed-down events for $D$ signals.

We use a Monte Carlo simulation (MC) of the production and decay
of signal and background events based on \evtgen~\cite{evtgen}.
A sample of simulated inclusive \BB events equivalent to about five times the integrated
luminosity is used to study backgrounds and to optimize event
selection criteria. Large samples of many individual semileptonic $B$ decays
(discussed in Section~\ref{sec:models}) are used to parameterize the
distributions of variables used in the fit. Final-state radiation is simulated using
{\tt PHOTOS}~\cite{photos}.
Simulation of the detector response is performed with
\geant~\cite{geant}
and the resulting efficiencies and resolutions are validated in
multiple data control samples.

\section{Semileptonic Decay Models}\label{sec:models}
In the SM, the matrix element for a semileptonic $B$
meson decay can be written as

\begin{equation}
\mathcal M\left(B\to\ds(\ellm/\taum)\nub\right)=-i\frac{g^2}{8m_W^2}V_{cb}L^\mu H_\mu~,
\end{equation}

\noindent where $g$ is the weak coupling constant,
$m_W$ the $W$ mass, $V_{cb}$ the quark mixing matrix element,
and $L^\mu$ and $H_\mu$ are the leptonic and hadronic currents,
respectively. Here, we have used a simplified form for the $W$
propagator appropriate for energies much less than $m_W$.
The leptonic current is exactly known,

\begin{equation}
L^\mu = \ubar_\ell \gamma^\mu(1-\gamma_5)v_\nu~,
\end{equation}

\noindent and the hadronic current is given by

\begin{equation}
H_\mu = \langle \ds | \cbar\gamma_\mu (1-\gamma_5)b|B\rangle~.
\end{equation}

In the case of a $B\to D$ transition, the axial-vector part
of the current does not contribute to the decay, and we may write the hadronic
current in terms of two form factors $f_+(q^2)$ and $f_-(q^2)$:

\begin{equation}
\langle D|V^\mu|B\rangle = (p+p')^\mu f_+(q^2)+(p-p')^\mu f_-(q^2)~,
\end{equation}

\noindent with $V^\mu\equiv\cbar\gamma^\mu b$ and where $p$
and $p'$ are the four-momenta of the $B$ and $D$ mesons,
respectively. For the $B\to\Dstar$ transition, the axial-vector
term contributes to the decay as well, and we write the hadronic current
in terms of form factors $V(q^2)$, $A_1(q^2)$, $A_2(q^2)$, $A_3(q^2)$, and $A_0(q^2)$:

\begin{multline}
\langle\Dstar|V^\mu-A^\mu|B\rangle =
  \frac{2i\epsilon^{\mu\nu\alpha\beta}}{m_B+m_{\Dstar}}\varepsilon^*_\nu p'_\alpha p_\beta V(q^2) - \\
 (m_B+m_{\Dstar})\varepsilon^{*\mu}A_1(q^2) + \frac{\varepsilon^*\cdot q}{m_B+m_{\Dstar}}(p+p')^\mu A_2(q^2) + \\
 2m_{\Dstar}\frac{\varepsilon^*\cdot q}{q^2}q^\mu A_3(q^2) - 2m_{\Dstar}\frac{\varepsilon^*\cdot q}{q^2}q^\mu A_0(q^2)~,
\end{multline}

\noindent where $A^\mu\equiv\cbar\gamma^\mu\gamma_5 b$ and
$\varepsilon$ is the \Dstar polarization vector. The form factor
$A_3(q^2)$ is related to two other form factors as

\begin{equation}
A_3(q^2)=\frac{m_B+m_{\Dstar}}{2m_{\Dstar}}A_1(q^2)-\frac{m_B-m_{\Dstar}}{2m_{\Dstar}}A_2(q^2)
\end{equation}

\noindent so that there are only four independent form factors.

In the limit of massless leptons, any terms proportional
to $q^\mu\equiv(p-p')^\mu$ vanish when the hadronic current is
contracted with the leptonic current. For this reason,
the contributions from the form factors $f_-(q^2)$ and
$A_0(q^2)$ are essentially negligible for electrons
and muons, as mentioned above.

Semileptonic decays are simulated using
the ISGW2 model~\cite{isgw2}, except for $B\to\Dstar\ellm\nulb$ decays,
which use an HQET model with a linear form factor expansion~\cite{hqet_linear}, and nonresonant
$B\to\ds\pi\ellm\nulb$ decays, which use the model of Goity and
Roberts~\cite{goityroberts}. We reweight both signal $B\to\ds\taum\nutb$
and normalization $B\to\ds\ellm\nulb$ events~\cite{ffreweight} so
that the decay distributions follow the Caprini-Lellouch-Neubert (CLN)
form factor model~\cite{CLN} with parameters measured in data.
We use $\rho_+^2=1.17\pm 0.18$~\cite{hfag} for $B\to D\ellm\nulb$
and $B\to D\taum\nutb$ decays, and we use
$R_1=1.417\pm 0.061\pm 0.044$, $R_2=0.836\pm 0.037\pm 0.022$, and
$\rho_{A_1}^2=1.179\pm 0.048\pm 0.028$~\cite{babarFF} for
$B\to\Dstar\ellm\nulb$ and $B\to\Dstar\taum\nutb$ decays.\footnote{The
parameters $R_1$ and $R_2$ are not included in the model of
Caprini, Lellouch, and Neubert~\cite{CLN}; to model the $B\to\Dstar$ form factors, we adopt
the formalism used in~\cite{babarFF2}, Eqs.~(13--14), where the leading terms
in these form factor ratio expansions are taken as free parameters.
We use independent slope parameters $\rho_+^2$ and $\rho_{A_1}^2$
for the $B\to D$ and $B\to\Dstar$ form factors, respectively,
treating the two sets of form factors as uncorrelated.}
Variation of these form factors
is taken into account as a systematic uncertainty,
including the correlations between the three
$B\to\Dstar$ form factor parameters.

\begin{figure}
\includegraphics[width=3.1in]{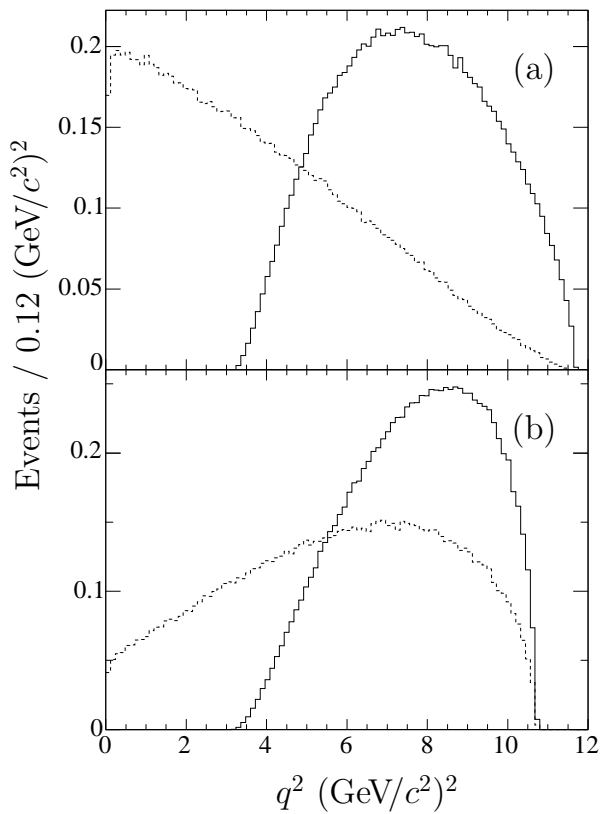}
\caption{Generated $q^2$ distributions for (a) $B\to D\ellm\nulb$ and $B\to D\taum\nutb$;
(b) $B\to\Dstar\ellm\nulb$ and $B\to\Dstar\taum\nutb$.
The two curves in each plot show $q^2$ for the light lepton (dashed) and for the $\tau$ (solid).
All distributions use the CLN form factor model with experimentally-measured shape parameters.
The distributions are normalized to equal areas.}
\label{fig:cln_q2}
\end{figure}

\begin{figure}
\includegraphics[width=3.1in]{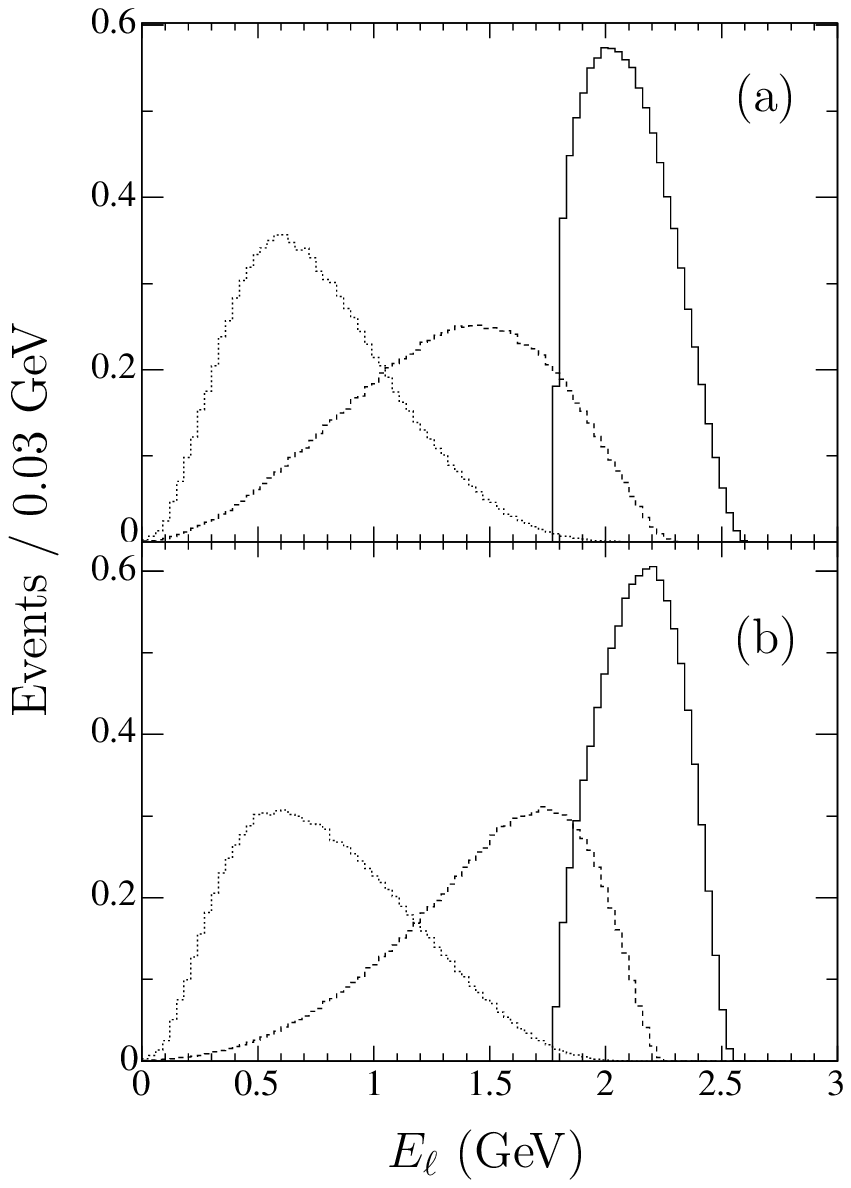}
\caption{Generated lepton energy distributions for (a) $B\to D\ellm\nulb$ and $B\to D\taum\nutb$;
(b) $B\to\Dstar\ellm\nulb$ and $B\to\Dstar\taum\nutb$.
The three curves in each plot show the \ellm energy in $B\to\ds\ellm\nulb$ (dashed),
the \taum energy in $B\to\ds\taum\nutb$ (solid), and the secondary
lepton energy in $B\to\ds\taum\nutb$ (dotted), all defined
in the $B$ meson rest frame.
All distributions use the CLN form factor model with experimentally-measured shape parameters.
The distributions are normalized to equal areas.}
\label{fig:cln_el}
\end{figure}

\begin{figure}
\includegraphics[width=3.1in]{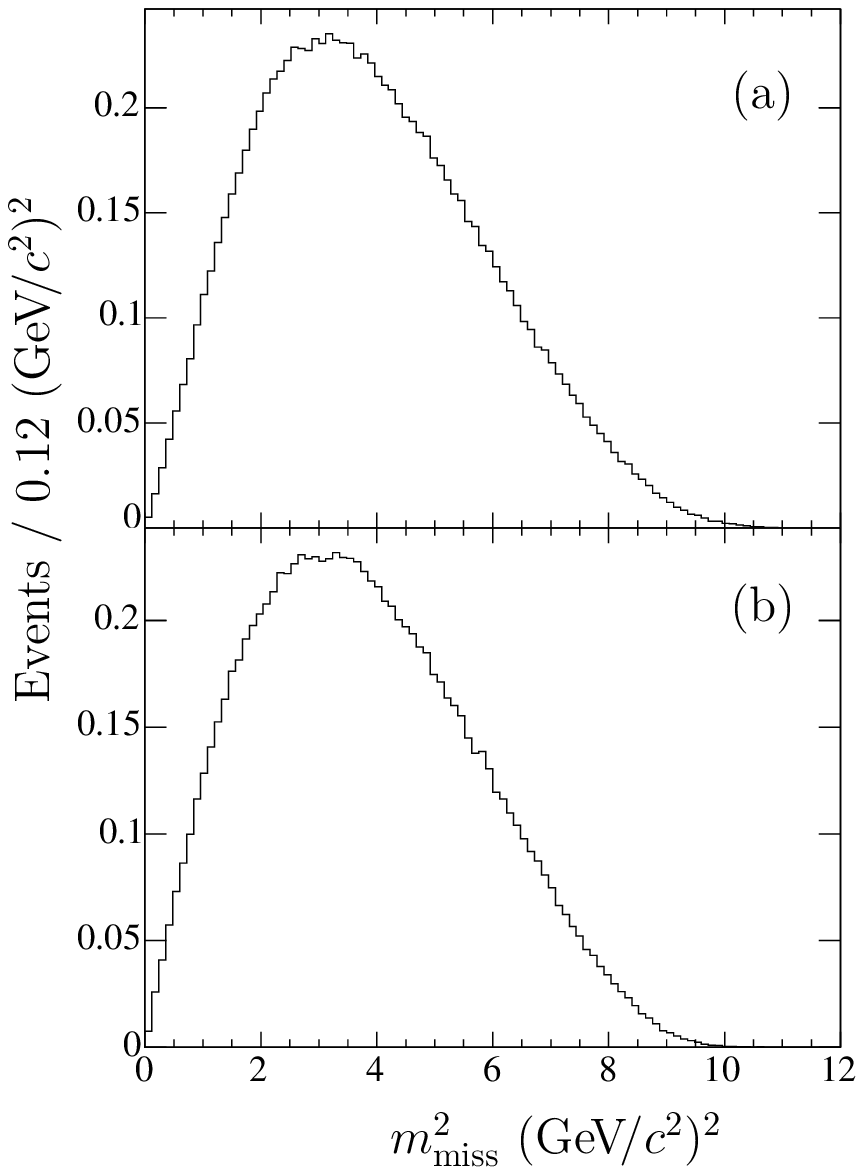}
\caption{Generated \mmiss distributions for (a) $B\to D\taum\nutb$ and (b) $B\to\Dstar\taum\nutb$.
Both distributions use the CLN form factor model with experimentally-measured shape parameters.}
\label{fig:cln_mm}
\end{figure}

Figures~\ref{fig:cln_q2}--\ref{fig:cln_mm} show distributions of
three kinematic variables important to this analysis, all generated
using the CLN form factor parameterization with parameters given
above. Figure~\ref{fig:cln_q2} compares $q^2$ distributions between
the signal and normalization modes. Signal events must
satisfy $q^2>m_\tau^2$, leading to qualitatively different $q^2$ spectra
for signal and normalization events; this feature is exploited
in the event selection and in validation studies. Figure~\ref{fig:cln_el}
shows distributions of lepton energy in the $B$ meson rest frame.
While the $\taum$ lepton in signal events typically has high energy (due to its mass),
the secondary lepton $\ellm$ typically has much lower energy than either
the $\taum$ or the primary lepton in $B\to\ds\ellm\nulb$ events. This
low lepton energy leads
to a lower reconstruction efficiency for signal leptons than those
in the normalization modes. Figure~\ref{fig:cln_mm} shows distributions
of \mmiss for the two signal modes, which, due to the three neutrinos
in these events, forms a broad structure up to very large \mmiss.

\section{Event Reconstruction and Selection}\label{sec:evtsel}
All event selection requirements (as well as the fit procedure
described in Section~\ref{sec:fit})
are defined using simulated events or using control samples
in data that exclude the signal region in order to avoid any
potential sources of bias. About 60\% of the \BB MC sample is used
in optimizing the event selection, while the remaining 40\%
is used as an independent validation of the selection and fitting procedures.

Most of the selection criteria described
here are optimized to maximize the quantity $S/\sqrt{S+B}$, where
$S$ and $B$ are the expected signal and background yields in the large
\mmiss region of our data
sample, assuming Standard Model branching fractions for signal
decays. The requirement on \DeltaE of the \btag candidate (defined below)
was initially optimized in the same way, but was tightened because fits
to MC samples indicated that events at large $|\DeltaE|$ contributed to biases
in the signal extraction. The final selection corresponds to a compromise
between the statistical $S/\sqrt{S+B}$ optimization and the systematic
effects due to this bias.

\subsection{\btag Reconstruction}
We reconstruct \btag candidates in 1114 final states
$\btag\to\ds Y^\pm$ with an algorithm that has been used previously
at \babar\ for a number of analyses, especially those dependent
on measuring missing momentum~\cite{BtagAlgorithm}.
These final states arise from the large number of ways to reconstruct the $D$
and \Dstar mesons within the \btag candidate and the possible pion and kaon
combinations within the $Y^\pm$ system.
Tag-side $D$ candidates are reconstructed as
$\Dz_\mathrm{tag}\to\Km\pip$, $\Km\pip\piz$, $\Km\pip\pipi$, and
$\KS\pipi$, and as $\Dp_\mathrm{tag}\to\Km\pip\pip$, $\Km\pip\pip\piz$,
$\KS\pip$, $\KS\pipi\pip$, and $\KS\pip\piz$. Tag-side \Dstar candidates
are reconstructed as $\Dstarz_\mathrm{tag}\to\Dz_\mathrm{tag}\piz$ and
$\Dz_\mathrm{tag}\gamma$ and as $\Dstarp_\mathrm{tag}\to\Dz_\mathrm{tag}\pip$.
The $Y^\pm$ system may consist
of up to six light hadrons (\pipm, \piz, \Kpm, or \KS).
In both the $\ds_\mathrm{tag}$ and $Y^\pm$ systems,
we reconstruct $\piz\to\gamma\gamma$ and $\KS\to\pipi$ and
require charged kaon candidates to satisfy PID criteria
(loose criteria for $\Dz\to\Km\pip$, tight
for all other modes.\footnote{The terms ``loose'' and ``tight'' refer to the
  relative signal-to-background discrimination of various PID
  criteria. Loose criteria are chosen to have high efficiency,
  and have relatively high background rates as well; tight
  criteria have lower background but also glower signal efficiency.
  The optimal choice of criteria depends on the particle
  type and on the {\it a priori} purity of the sample, and
  is therefore different for each reconstruction channel.})
$\ds_\mathrm{tag}$ candidates are
selected within about $2\sigma$ (standard deviations)
of the nominal mass, with $\sigma$
depending on the reconstruction mode and typically
$5$--$10\ \mevcc$ for the $D_\mathrm{tag}$ mass and $1$--$2\ \mevcc$ for the $\Dstar_\mathrm{tag}-D_\mathrm{tag}$ mass difference.

We use two kinematic variables to identify \btag candidates,

\begin{equation}
\mes=\sqrt{s/4-|{\bf p}_\mathrm{tag}|^2}
\end{equation}

\noindent and

\begin{equation}
\DeltaE=E_\mathrm{tag}-\sqrt{s}/2~,
\end{equation}

\noindent where $\sqrt{s}$ is the total \epem
energy, $|{\bf p}_\mathrm{tag}|$ is the magnitude of the \btag momentum, and $E_\mathrm{tag}$ is
the \btag energy, all defined in the \epem center-of-mass frame.
For correctly reconstructed \btag candidates, \mes is equal
to the $B$ meson mass, with a resolution of about $2.5\ \mevcc$, and
\DeltaE is equal to zero, with a resolution of about $18\ \mev$.

For each $\ds_\mathrm{tag}$ ``seed'' candidate, we 
use a recursive algorithm to identify candidate $Y^\pm$ systems.
Light hadrons from the remaining tracks and photons in the event
are added to the $Y^\pm$ system, one at a time. If the
resulting values of \mes and \DeltaE for the $\ds_\mathrm{tag}Y^\pm$
candidate are close to the nominal values, the \btag candidate
is accepted. If the value of \DeltaE is too large, the light
hadron just added is removed from the $Y^\pm$ system, since
continuing to add particles to this $Y^\pm$ candidate
will increase \DeltaE further. The algorithm then continues
recursively with the remaining particles in the event,
adding and removing light hadrons to the $Y^\pm$ system
according to \mes, \DeltaE, and the $Y^\pm$ system topology.
This algorithm is semiexclusive, meaning that particles in
the $Y^\pm$ system are not constrained to intermediate resonance
states. Because of this, the yield is significantly higher than
exclusive $B$ reconstruction, while the purity is somewhat lower.
In this analysis, however, since we exclusively reconstruct the second $B$ meson in the
event, the purity of our final sample is substantially improved
with respect to the raw \btag sample.

\begin{figure}
\includegraphics[width=3.1in]{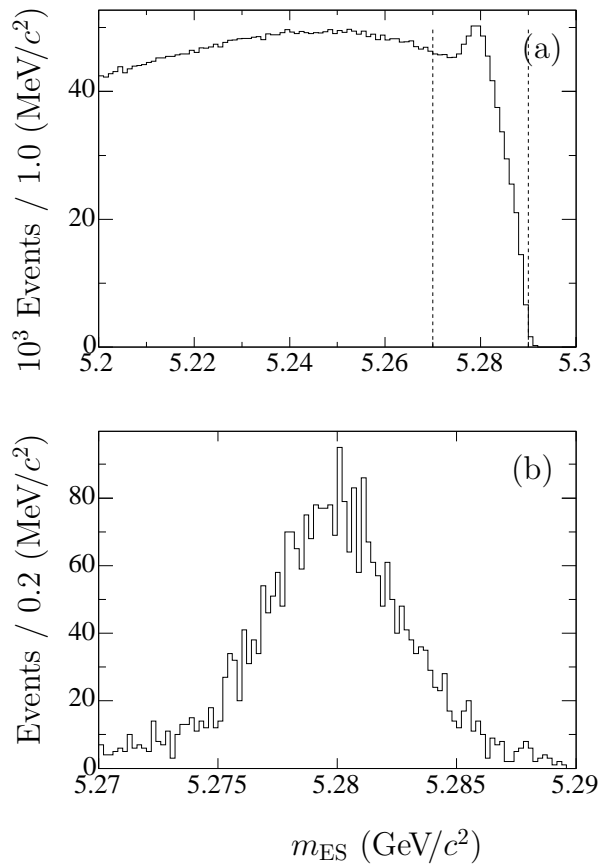}
\caption{Distributions of \mes in selected events. (a)
shows all \btag candidates reconstructed in $20\ \invfb$
of data, and the purity of this plot has been increased
by requiring $|\DeltaE|<50\ \mev$. (b) shows the
distribution for the complete data sample after
the signal $B$ and total-event selection requirements.
Note the substantial improvement in purity due to the
complete reconstruction.}
\label{fig:mes}
\end{figure}

We require $\mes>5.27\ \gevcc$ and $|\DeltaE|<72\ \mev$,
corresponding to $\pm 4\sigma$ in
\DeltaE and $-4\sigma$ in \mes (the kinematic limit $\mes<\sqrt{s}/2$ provides an effective $+4\sigma$ requirement). We reconstruct \btag candidates
with an efficiency of $0.2\%$ to $0.3\%$.
Figure~\ref{fig:mes} shows
distributions of \mes for selected \btag candidates both
before and after the signal-side reconstruction.
We make no attempt at this stage
to select a single \btag among multiple reconstructed
candidates: this decision is made after reconstructing
the signal side as well.

\subsection{Reconstruction of the Signal $B$}
For the $B$ meson decaying semileptonically, we reconstruct \ds
candidates in the modes $\Dz\to\Km\pip$,
$\Km\pip\piz$, $\Km\pip\pipi$, $\KS\pipi$; $\Dp\to\Km\pip\pip$,
$\Km\pip\pip\piz$, $\KS\pip$, $\Km\Kp\pip$; $\Dstarz\to\Dz\piz$, $\Dz\gamma$;
and $\Dstarp\to\Dz\pip$, $\Dp\piz$. We reconstruct \KS mesons
as $\KS\to\pipi$ with $491<m_{\pipi}<506\mevcc$, corresponding
to $\pm 3\sigma$. We reconstruct \piz mesons as $\piz\to\gamma\gamma$,
requiring $90<m_{\gamma\gamma}<165\mevcc$ for the soft \piz used
to reconstruct $\Dstar\to D\pi^0_\mathrm{soft}$, and
requiring $E_{\gamma\gamma}>200\mev$ and $125<m_{\gamma\gamma}<145\mevcc$
for a \piz used to reconstruct a $D$ meson; the mass intervals correspond
to $\pm 3\sigma$ in both cases and are different because the
resolution is poorer at low energies. Charged kaon candidates are
required to satisfy tight PID criteria with a
typical efficiency of 85\% while rejecting 98\% of pions.
Charged pion candidates are required to satisfy loose
PID criteria with a typical efficiency of 97\% while
rejecting 88\% of kaons.
$D$ (\Dstar) candidates are selected within
$4\sigma$ of the $D$ mass ($\Dstar-D$ mass difference);
as on the tag side, $\sigma$ is typically
$5$--$10\ \mevcc$ ($1$--$2\ \mevcc$).

Electron candidates are required to satisfy tight PID criteria and
to have lab-frame momentum $|{\bf p}_e|>300 \mevc$, with an
efficiency that rises from 85\% at the lowest momenta to 95\%
for $|{\bf p}_e|> 1.0\ \gevc$. Muon candidates are required
to satisfy tight PID criteria; since muon PID relies on the hit
pattern in the IFR, this effectively requires
$|{\bf p}_{\mu}|\gtrapprox 600 \mevc$, and
results in an efficiency of 40\%--60\% over the
allowed momentum range.
The energy of electron candidates is corrected for bremsstrahlung energy loss
if photons are found close to the electron direction. Lepton candidates
of either flavor are required to have at least 12 hits in the drift chamber
and to have a laboratory-frame polar angle $0.4<\theta<2.6\ \mbox{rad}$
(excluding the very forward and very backward regions of the
tracking system) in order to ensure a well-measured momentum, since mismeasured
lepton momenta distort the \mmiss distribution and tend to move
background events into the signal-like region. Approximately
5\% of selected lepton candidates are misidentified, almost
all of which are pions misreconstructed as muons.

\subsection{Total-Event and Single-Candidate Selection}
We form whole-event candidates by combining \btag candidates
with $\ds\ellm$ candidate systems. We combine charged
\btag candidates with $D^{(*)0}\ellm$ systems
and neutral \btag candidates with both $D^{(*)+}\ellm$
and $D^{(*)-}\ellp$ systems, where the inclusion
of both charge combinations allows for neutral $B$ mixing.

\begin{figure}
\includegraphics[width=3.1in]{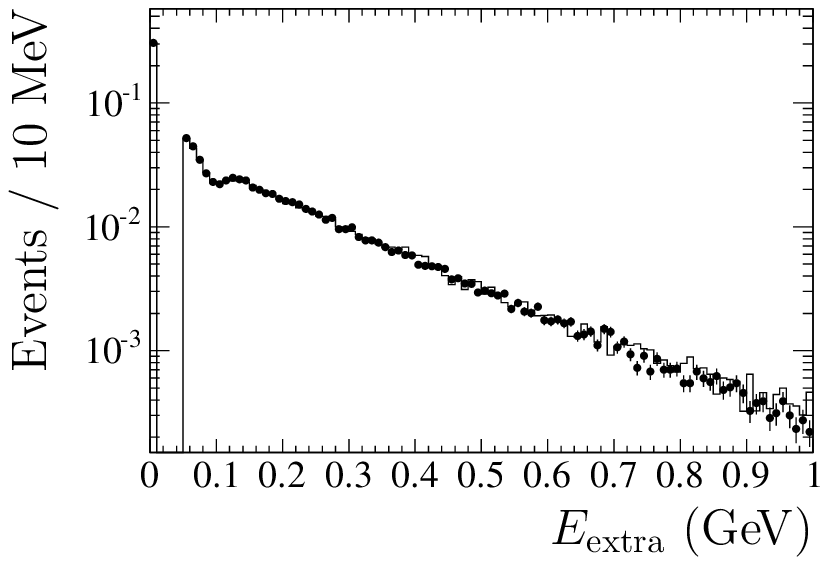}
\caption{Distributions of \eextra in simulated events, shown after all
event selection except the cut on \eextra itself. Signal $B\to\ds\taum\nutb$
events are shown as points, while $B\to\ds\ellm\nulb$ normalization events
are shown as the histogram. The gap between the $\eextra=0$ bin and the
remainder of the distribution corresponds to the minimum allowed
photon energy, $50\ \mev$. The normalization is arbitrary. The
agreement between the two distributions indicates that the efficiency
of a cut on \eextra will cancel when we measure the branching-fraction ratio.}
\label{fig:eecomp}
\end{figure}

In correctly reconstructed signal and normalization events,
all of the stable final-state particles, with the exception
of the neutrinos, are associated with either the \btag,
\ds, or $\ellm$ candidate. Events with additional particles
in the final state must therefore have been misreconstructed,
and we suppress these backgrounds with two selection requirements
on the ``extra'' particle content in the event.
We require that all observed charged tracks be associated with either
the \btag, \ds, or $\ell$ candidate.
We compute \eextra, the sum of the
energies of all photon candidates not associated with the $\btag+\ds\ell$ 
candidate system, and we require $\eextra<150$--$300\ \mev$, depending
on the \ds channel.
When considering these extra tracks and extra photons,
care is taken to reject track and photon candidates which
are likely to be due to accelerator background,
electronics noise, or reconstruction software failures;
fake photons in the EMC are, to some degree,
unavoidable, which is why we can not simply require $\eextra=0$.
The different $D$ modes have very different
levels of combinatorial background, which the \eextra cut
is particularly effective at rejecting.
Figure~\ref{fig:eecomp} shows distributions of \eextra
for simulated signal and normalization events. Excellent
agreement is seen in the two distributions, indicating that
the efficiency of a cut on \eextra will largely cancel when we
measure the branching-fraction ratio; we observe the same level of
agreement in the four $\ds\ellm$ channels separately, as well
as in the $e$ and $\mu$ final states separately.

We suppress hadronic events and combinatorial
backgrounds by requiring $|{\bf p}_\mathrm{miss}|>200\ \mevc$,
where $|{\bf p}_\mathrm{miss}|$ is the magnitude of the
missing momentum. This requirement
mainly rejects hadronic events such as $B\to\ds\pim$, where the
\pim is misidentified as a \mun. Our selection rejects more than
99\% of $B\to\ds\pim$ background, while rejecting less than
1\% of signal and other semileptonic events. 

We further suppress background by requiring $q^2>4\ (\gevccnosp)^2$,
where $q^2$ is calculated as 

\begin{equation}
q^2=[p_B-p_{\ds}]^2=[p_{\epem}-p_{\rm tag}-p_{\ds}]^2~.
\label{eq:q2def}
\end{equation}

\noindent This requirement
preferentially rejects combinatorial backgrounds from two-body
$B$ decays such as $B\to \ds D$, where one $D$ meson decays
semileptonically (or, in the case of a \Ds, leptonically as
$\Ds\to\taup\nut$). Our selection rejects about 25\% of
these backgrounds, while the signal efficiency is about
98\% because signal events automatically satisfy
$q^2>m_\tau^2\approx 3.16\ (\gevccnosp)^2$.
For $B\to D\ellm\nulb$ decays, the $q^2$
distribution peaks near zero (see Fig.~\ref{fig:cln_q2}),
so this selection has an efficiency of about 60\% for this normalization mode; for
$B\to\Dstar\ellm\nulb$ decays, the $q^2$ distribution
peaks at higher values, so our efficiency is
about 70\%. The $q^2$ requirement
is the main reason why the reconstruction efficiency
is different for signal and normalization modes, as
seen below.

If multiple candidate systems pass our selection in a given event,
we select the one with the lowest value of \eextra. This scheme preferentially
selects the candidate that is least likely to have lost additional particles.
The main effect of this algorithm is that a candidate in one
of the $\Dstar\ellm$ channels will be selected before a candidate in one
of the $D\ellm$ channels when both candidates are present in an event.
Because $\Dstar\to D$ feed-down is a dominant background while
$D\to\Dstar$ feed-up is comparatively rare, keeping as many
true \Dstar events in the $\Dstar\ellm$ reconstruction channels helps to
increase the sensitivity to the $D\taum\nutb$ signals.

To improve the resolution on the missing momentum, we perform a kinematic
fit~\cite{TreeFitter} to all $\FourS\to\btag\ds\ellm$ candidates.
We constrain charged track daughters of \KS, $D$, and $B$ mesons
to originate from common vertices, and we constrain the $\FourS\to\BB$
vertex to be consistent with the measured \babar\ beamspot location.
We constrain the mass of the signal $D$ meson (and \Dstar meson, if
there is one) to the measured value~\cite{PDG}, and
the combined momentum of the two $B$ mesons to be consistent with the measured
beam energy.

\subsection{$D^{**}$ Control Sample Selection}
We select four control samples
to constrain the poorly known $B\to D^{**}(\ellm/\taum)\nub$ background.
The selection is identical to that of the signal channels, but 
we require the presence of a \piz meson in addition to the $\btag+\ds\ell$
system. The \piz candidate must have momentum greater than $400\ \mevc$, and
the event must satisfy $\eextra<500\ \mev$, where the two photons
from $\piz\to\gamma\gamma$ are excluded from the calculation of
\eextra.

Most of the $D^{**}$ background in the four
signal channels occurs when the \piz from $D^{**}\to\ds\piz$ is not
reconstructed, so these control samples provide a direct normalization of
the background source. Similar $D^{**}$ decays in which a $\pipm$
is lost contribute very little to the background since they do not have
the correct charge correlation between the \btag and \ds candidate, and
decays with two missing charged pions, which may have the correct
charge correlation, have very low reconstruction efficiencies.
The feed-down probabilities for the $D^{**}(\ellm/\taum)\nub$
background are determined from simulation, with uncertainties in the
$D^{**}$ content treated as a systematic error as described in Sec.~\ref{sub:dss}.

\section{Selected Event Samples}
After applying all of the criteria above, we select a total of 3196 data
events, 2886 in the four signal channels and 310 in the
$D^{**}$ control samples, as listed in Table~\ref{tab:nevents}.
Since most of the events at large \mmiss are either $D\taum\nutb$
or $\Dstar\taum\nutb$ signal events, the third column
in the Table gives a first indication of where our sensitivity
comes from. There are more events in the two \Bm channels, $\Dz\ellm$
and $\Dstarz\ellm$, due to a larger efficiency to reconstruct
charged \btag candidates than neutral ones and, to a lesser extent,
a larger efficiency to reconstruct \Dz mesons on the signal side than
\Dp mesons. There are more events in the $D$ channels than the \Dstar
channels, particularly at large \mmiss, because these channels contain
both $D$ mesons and \Dstar feed-down. The greatest signal sensitivity
therefore comes from the $\Dz\ellm$ channel.

\begin{table}
\caption{Number of selected data events in the four signal channels, $N_\mathrm{ev}$,
and in the four $D^{**}$ control samples, $N_{D^{**}\mathrm{CS}}$.
Here, the large \mmiss region is taken to
be $\mmiss>2\ (\gevccnosp)^2$ and corresponds to the region with greatest
signal sensitivity.}
\label{tab:nevents}
\begin{tabular}{l l l l}\hline\hline
Channel & $N_\mathrm{ev}$\qquad\qquad & $N_\mathrm{ev}$ (large \mmiss) & $N_{D^{**}\mathrm{CS}}$\\ \hline
$\Dz\ellm$         & 1403 & 121 & 137\\
$\Dstarz\ellm$     & 790 & 43 & 77\\
$\Dp\ellm$         & 295 & 36 & 66\\
$\Dstarp\ellm$     & 398 & 14 & 30\\\hline\hline
\end{tabular}
\end{table}

\begin{figure}
\includegraphics[width=3.1in]{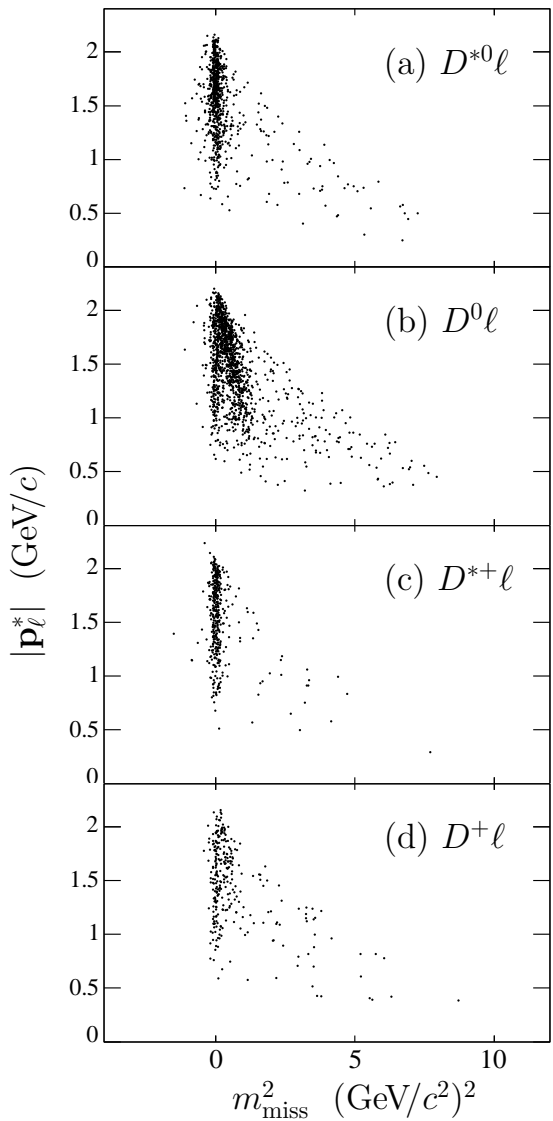}
\caption{Distributions of \pstarl versus \mmiss for selected
data events in the four signal channels.}
\label{fig:scatter}
\end{figure}

Figure~\ref{fig:scatter} shows distributions of \pstarl versus
\mmiss for the selected data samples. One-dimensional
distributions of \mmiss and \pstarl for these samples
are shown when we discuss the signal fit in Section~\ref{sec:fit}.

\begin{figure}
\includegraphics[width=3.1in]{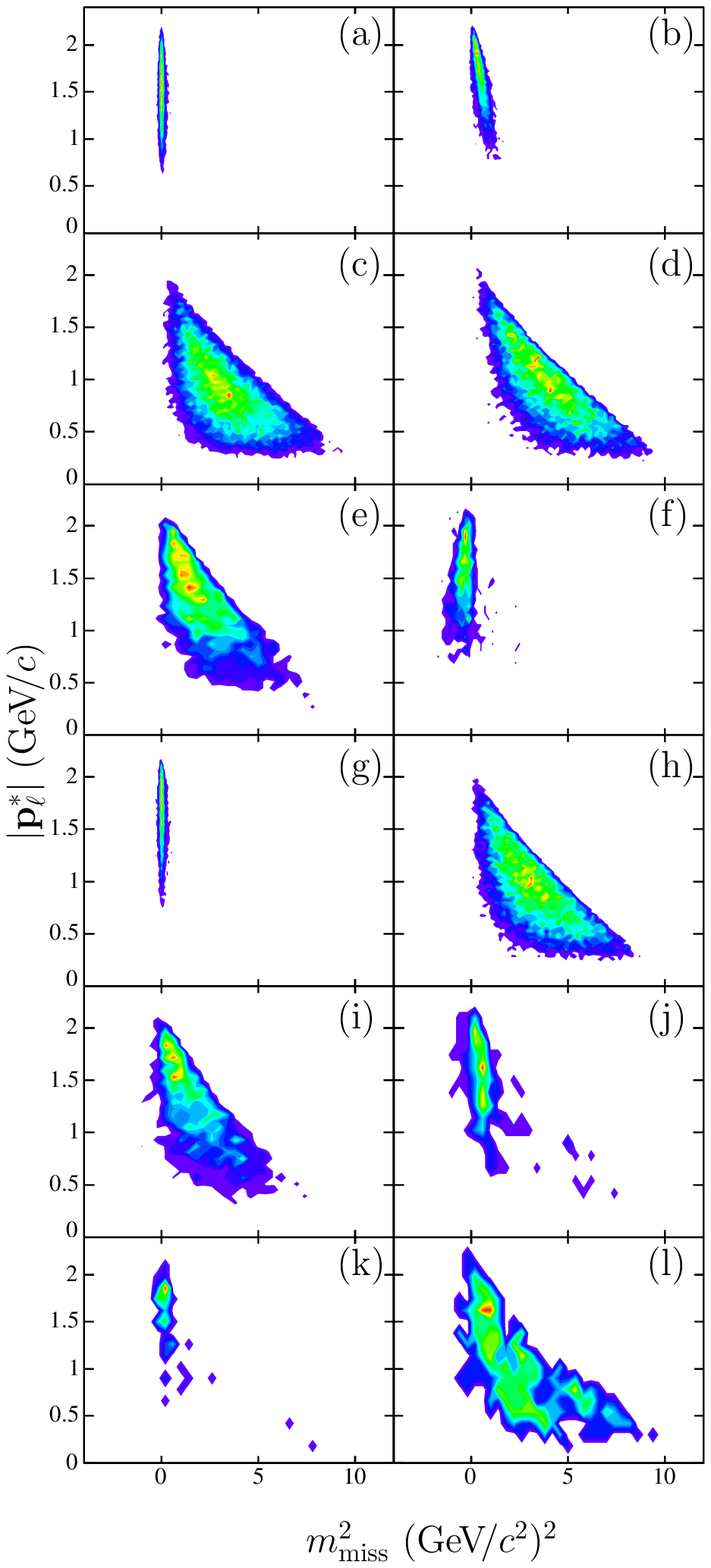}
\caption{(Color online) Distributions of \pstarl versus \mmiss for several MC samples
after all event selection. Red (light grey) regions indicate relatively
high density of reconstructed events, while blue (dark grey) indicate
relatively low density.
Shown are (a) $\Dz\ellm\nulb\Rightarrow\Dz\ellm$,
(b) $\Dstarz\ellm\nulb\Rightarrow\Dz\ellm$, (c) $\Dz\taum\nutb\Rightarrow\Dz\ellm$,
(d) $\Dstarz\taum\nutb\Rightarrow\Dz\ellm$, (e) $D^{**}\ellm\nulb\Rightarrow\Dz\ellm$,
(f) $\Dz\ellm\nulb\Rightarrow\Dstarz\ellm$, (g) $\Dstarz\ellm\nulb\Rightarrow\Dstarz\ellm$,
(h) $\Dstarz\taum\nutb\Rightarrow\Dstarz\ellm$, (i) $D^{**}\ellm\nulb\Rightarrow\Dstarz\ellm$,
(j) charge-crossfeed reconstructed in the $\Dz\ellm$ and $\Dp\ellm$ channels,
(k) charge-crossfeed reconstructed in the $\Dstarz\ellm$ and $\Dstarp\ellm$ channels,
and (l) combinatorial background in the four $\ds\ellm$ channels. The
reconstruction channel notation $\Rightarrow$ and the features of
these distributions are discussed in the text.}
\label{fig:mc2d}
\end{figure}

Figure~\ref{fig:mc2d} shows distributions of \pstarl versus \mmiss for
several MC samples after applying all event selection criteria. While the composition of the
event sample will be discussed in greater detail in the following section,
these distributions exhibit the qualitative features of the
data sample which are most relevant to our signal extraction.
Figure~\ref{fig:mc2d}(a) shows $\Dz\ellm\nulb\Rightarrow\Dz\ellm$,
where we introduce the $\Rightarrow$ notation to mean that these
are true $\Bm\to\Dz\ellm\nulb$ events reconstructed in
the $\Dz\ellm$ channel. The \mmiss distribution is very narrowly
peaked around zero, as expected for one-neutrino events.
Figure~\ref{fig:mc2d}(b) shows $\Dstarz\ellm\nulb\Rightarrow\Dz\ellm$,
feed-down events where a \Dstarz is misreconstructed as a
\Dz. In this case, the center of the \mmiss distribution is
offset from zero, and this offset decreases with increasing
\pstarl; this kinematic feature is common to all feed-down
processes, and is due to the fact that higher \pstarl correspond
to lower \Dstar momenta and therefore to lower momenta for
the lost \piz or $\gamma$. The width of the \mmiss distribution
is also observed to decrease with increasing \pstarl, a feature
which is also common to most distributions; this narrowing is
partly due to the same kinematic effect as before, the reduced
\Dstar phase space at large \pstarl, and partly due to the fact
that the lepton momentum resolution improves at higher momenta.

Figures~\ref{fig:mc2d}(c) and (d) show similar distributions for
signal events, (c) showing correctly reconstructed $\Bm\to\Dz\taum\nutb$
and (d) showing \Dstarz feed-down; in both plots, the large
values of \mmiss due to the three neutrinos are clearly observed. Again,
the \mmiss distributions move towards zero and become narrower
at high \pstarl, in this case due to the reduced phase space
for the multiple neutrinos, although, in Fig.~\ref{fig:mc2d}(d),
the effect of the lost \piz or $\gamma$ can also be seen as a defecit
along the lower-left edge of the distribution.
Figure~\ref{fig:mc2d}(e) shows feed-down from $B\to D^{**}\ellm\nulb$
into the $\Dz\ellm$ channel, where, in addition to the neutrino,
one or more \piz mesons or photons from the $D^{**}$ decay have
been lost. Since \piz mesons from $D^{**}$ decay typically have
higher momentum than those from \Dstar decay, the \mmiss distribution
is much broader than that in Fig.~\ref{fig:mc2d}(b).
Figure~\ref{fig:mc2d}(f) shows the feed-up process
$\Dz\ellm\nulb\Rightarrow\Dstarz\ellm$, where a true \Dz meson is
paired with a combinatorial \piz or $\gamma$ to fake a \Dstarz
candidate. In this case, the \mmiss distribution is shifted in
the opposite direction from Fig.~\ref{fig:mc2d}(b).
Figures~\ref{fig:mc2d}(g), (h), and (i) show three additional
distributions for events reconstructed in the $\Dstarz\ellm$
channel, $\Dstarz\ellm\nulb$, $\Dstarz\taum\nutb$ signal,
and $D^{**}\ellm\nulb$ background, respectively; each of these
distributions is similar to the corresponding one in the
$\Dz\ellm$ channel, Figs.~\ref{fig:mc2d}(a), (c), and (e),
respectively.

Figures~\ref{fig:mc2d}(j) and (k) show charge-crossfeed backgrounds:
true $B\to\ds\ellm\nulb$ events reconstructed with the wrong
charge for both the \btag and \ds meson. Typically this occurs
when a low-momentum \pipm is swapped between the two mesons.
Note that, even though the event is misreconstructed, this
particle misassignment does not substantially alter the total missing
momentum, so that the \mmiss distribution still peaks at or near zero.
While the events in Fig.~\ref{fig:mc2d}(k), which are
reconstructed in the $\Dstar\ellm$ channels, are very strongly
peaked at $\mmiss=0$, Fig.~\ref{fig:mc2d}(j) includes a large
feed-down component, and therefore exhibits the same sloping behavior
seen in Fig.~\ref{fig:mc2d}(b).

Figure~\ref{fig:mc2d}(l) shows the distribution for combinatorial
background for all four signal channels. This background is
dominated by hadronic $B$ decays such as $B\to\ds D_s^{(*)}$
that produce a secondary lepton, including events with $\tau$ leptons from
$D_s$ decay.

In our \BB MC sample, our criteria select $D^{**}$ control samples which
are $60\%$--$80\%$ pure $B\to D^{**}\ellm\nulb$ events, of which
more than $90\%$ involve true $D^{**}\to\ds\piz$ transitions.
The remaining events are split between feed-up from
$B\to\ds\ellm\nulb$ and combinatorial background. In these
control samples, the $B\to D^{**}\ellm\nulb$ component peaks
at or near zero in \mmiss, just as $B\to\ds\ellm\nulb$ does
in the four signal channels. The qualitative features of the
other contributions are similar to what is seen in the signal
channels.

\section{Kinematic Control Samples}
The event selection criteria described in Section~\ref{sec:evtsel}
are more complicated than those used in a typical \babar\ analysis, due to the
full-event reconstruction of a high-multiplicity final state and the
need to veto events with extra tracks and neutral clusters. We use two
data control samples to validate our simulation with respect to the observed
behavior in data. The control samples are kinematically selected, with
no requirement on \mmiss, to be high purity samples of $B\to\ds\ellm\nulb$
events, with little or no contamination from signal decays.

The first control sample is defined by requiring the reconstructed lepton
to satisfy $\pstarl>1.5\ \gevc$, and is therefore a subset of our full
analysis sample. In simulation, 95\% of the selected sample is
$B\to\ds\ellm\nulb$ (the two $\Dstar\ellm$ channels are approximately 95\%
$B\to\Dstar\ellm\nulb$, while the two $D\ellm$ channels include both
$B\to D\ellm\nulb$ and large feed-down from $B\to\Dstar\ellm\nulb$). The
remaining 5\% of the sample is composed of about 1\%--3\% $B\to D^{**}\ellm\nulb$,
less than 1\% $B\to\ds\taum\nutb$, and 1\%--2\% of combinatorial backgrounds.

For the second control sample, we remove the standard $q^2>4\ (\gevccnosp)^2$
selection and require that events instead satisfy
$q^2<5\ (\gevccnosp)^2$, with $q^2$ calculated according to Eq.~(\ref{eq:q2def}). This control sample has very little overlap
with our final event sample, where we require $q^2>4\ (\gevccnosp)^2$.
Although the two control samples do have some overlap, this $q^2$ control sample has
the advantage over the first of allowing us to examine events with low \pstarl, as
expected for signal events. In simulation, approximately 90\% of this control
sample is $B\to\ds\ellm\nulb$ (as in the first sample, the two
$\Dstar\ellm$ channels are approximately 90\% $B\to\Dstar\ellm\nulb$, while
the two $D\ellm$ channels include \Dstar feed-down). The remainder of the
sample is composed of about 3\% $B\to D^{**}\ellm\nulb$, 3\%
$B\to\ds\taum\nutb$, and 4\%--5\% combinatorial backgrounds.

\begin{figure*}
\includegraphics[width=2.25in]{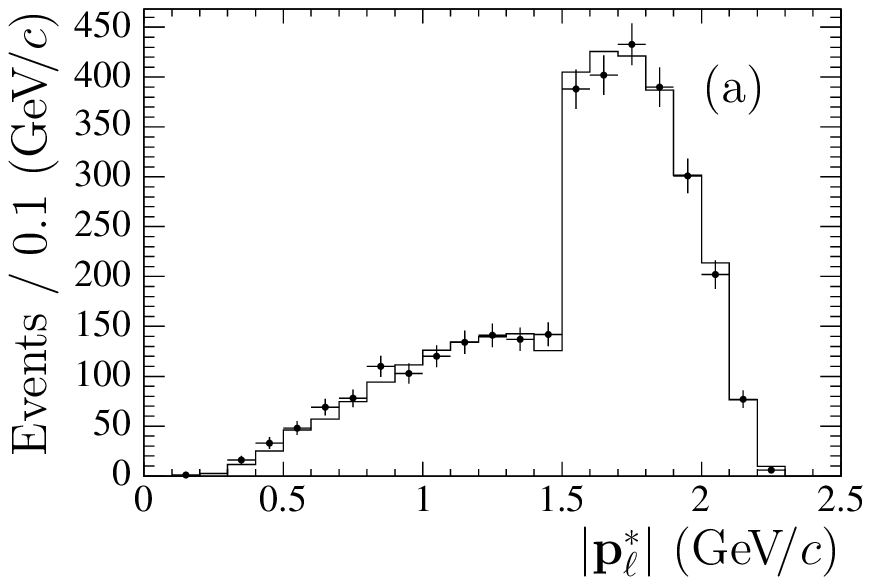}
\includegraphics[width=2.25in]{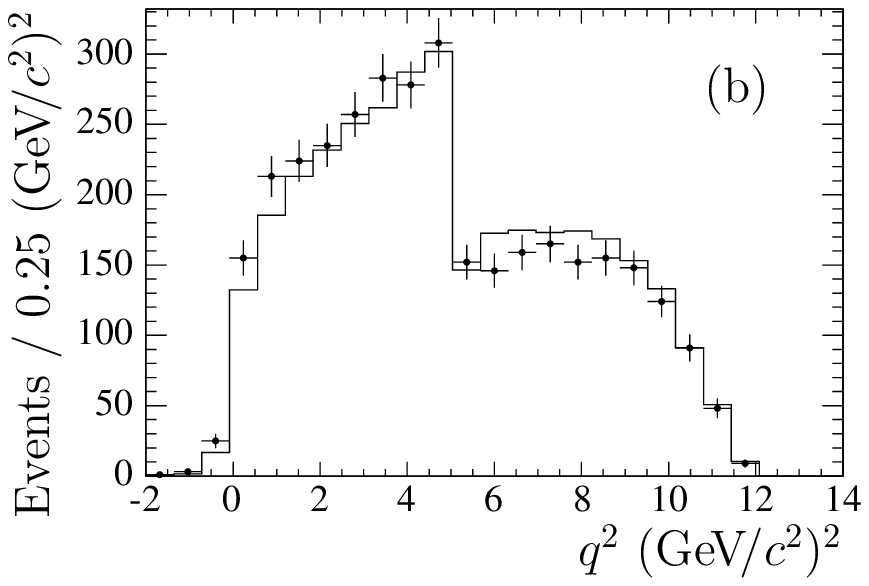}
\includegraphics[width=2.25in]{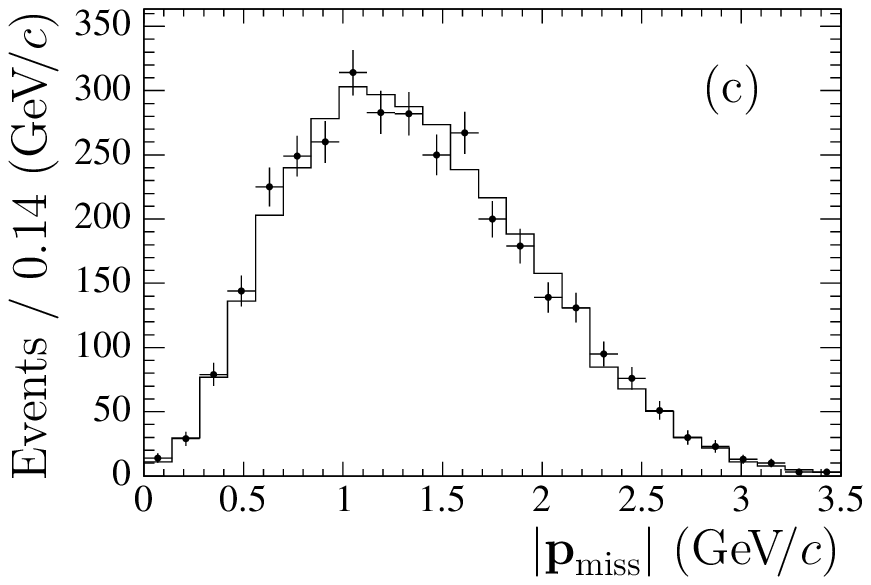}\\ \vspace{0.1in}
\includegraphics[width=2.25in]{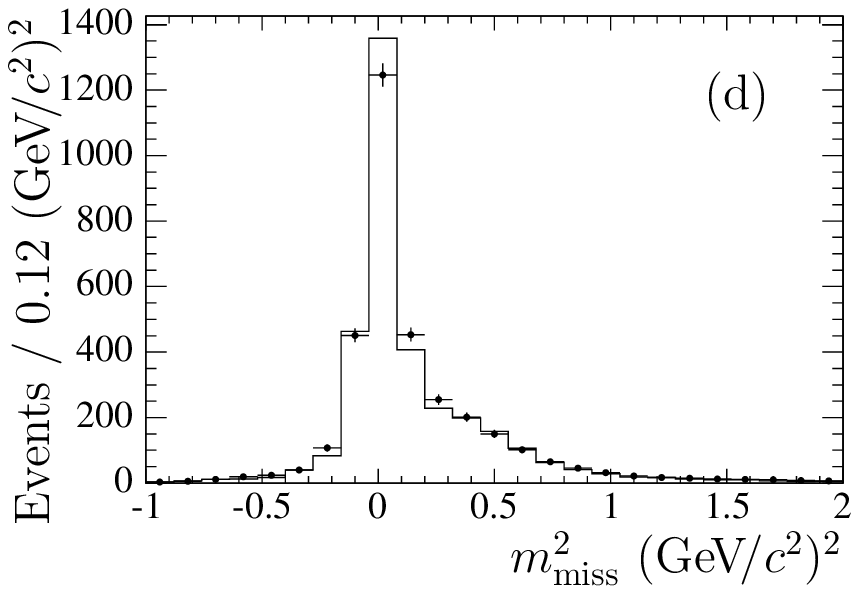}
\includegraphics[width=2.25in]{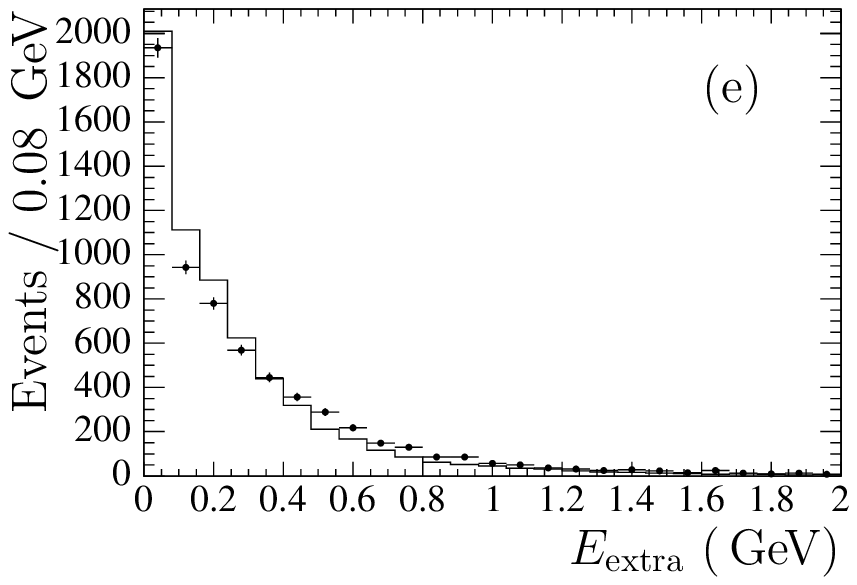}
\includegraphics[width=2.25in]{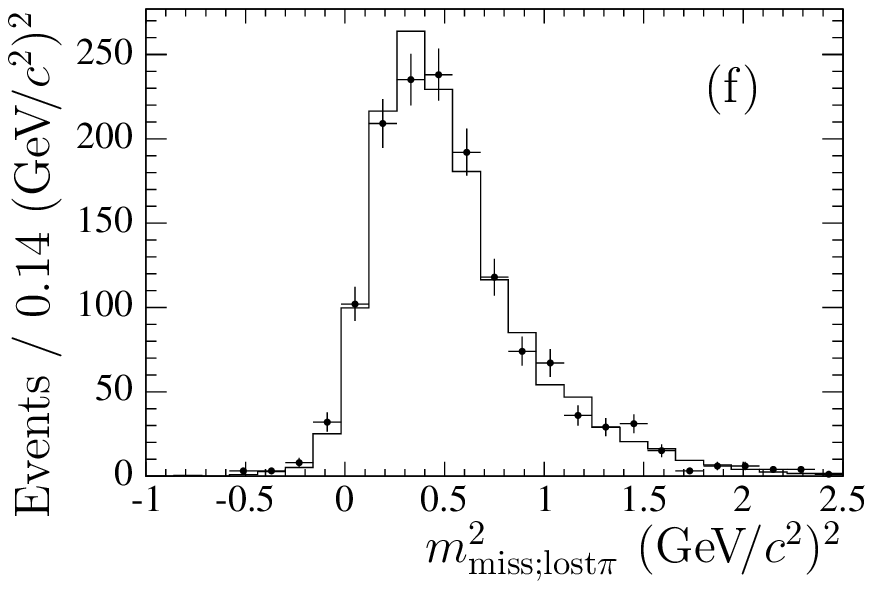}\\ \vspace{0.1in}
\includegraphics[width=2.25in]{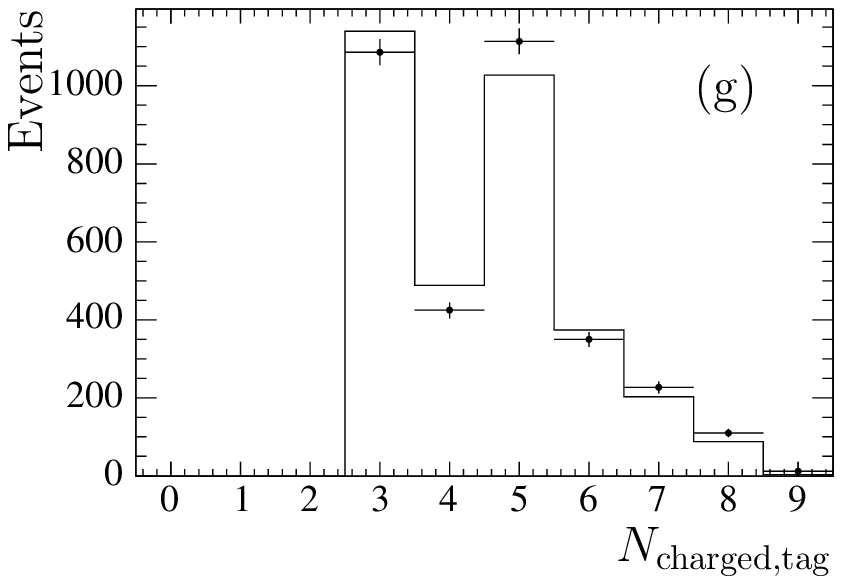}
\includegraphics[width=2.25in]{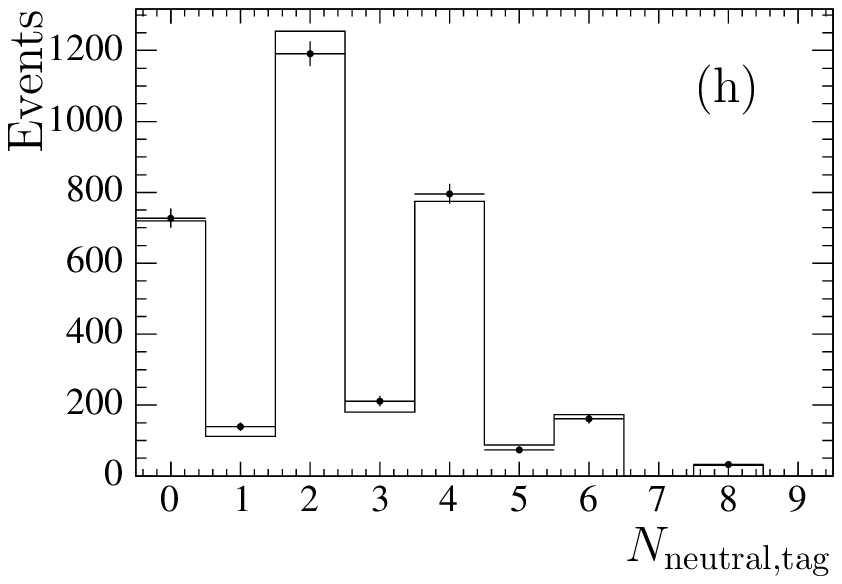}
\caption{Kinematic control sample plots: (a) \pstarl; (b) $q^2$; (c) $|{\bf p}_\mathrm{miss}|$;
(d) \mmiss; (e) \eextra; (f) \mmiss in the \Dstar channels, assuming that the
soft $\pi/\gamma$ had been lost; multiplicity of (g) charged tracks and (h) neutral
clusters used to reconstruct the \btag. In all plots, the points with error bars
are the data and the solid histogram is the simulation, scaled to the data luminosity.
Good agreement is seen between data and simulation in a variety of variables corresponding
to reconstruction, kinematics, and hermeticity requirements.
Small differences between data and simulation cancel in the relative
measurement and have no detrimental effect on the analysis.
The large steps in (a) and (b) are due to the combination of two control samples,
as described in the text.
The structure in (g) is caused by the larger efficiency to reconstruct
charged \btag candidates---with an odd number of charged tracks---than
neutral candidates, while the prominent even-odd structure in (h)
is due to the fact that most neutral clusters correspond to the
process $\piz\to\gamma\gamma$ and so appear in pairs.}
\label{fig:control}
\end{figure*}

Figure~\ref{fig:control} shows several data-simulation comparisons in the
two control samples. The four $\ds\ellm$ channels have been combined
in these plots as have the two control samples, and this union of the two control samples
is responsible for the large steps visible in (a) and (b). We see
good agreement between data and simulation in these plots, as
well as in similar studies where the two control samples are examined
separately, the four $\ds\ellm$ channels are examined separately, the two
lepton types are examined separately, and where the data are split
according to \babar\ running period. We have examined variables
related to \btag reconstruction, signal-side reconstruction,
hermeticity and whole-event reconstruction, and missing momentum.
In all cases, we observe that the simulation does a reasonable
job describing the data.
Because of the relative normalization scheme, small differences
between simulation and data have no detrimental effect on the
analysis.

\section{Fit of Signal and Background Yields}\label{sec:fit}
\subsection{Fit Overview}
Signal and background yields are extracted 
using an extended unbinned maximum likelihood fit to the 
joint (\mmiss, \pstarl) distribution. The fit
is performed simultaneously in the four signal channels and the four $D^{**}$
control samples. Two two-dimensional probability density
functions (PDFs) are presented in Section~\ref{sub:pdf};
each component in the fit (listed below) is described by
one of these two PDFs, with parameters determined from fits
to simulated event samples. A set of constraints, described
in Section~\ref{sub:constraints}, relate fit components in
different reconstruction channels. These constraints
are also determined from MC samples, except for parameters
describing the amount of \Dstar feed-down into the $D\ellm$
signal channels, which are determined directly by the
fit to data.

Tables~\ref{tab:comps1} and~\ref{tab:comps2} summarize the
parameterization of the fit in the four signal channels and the four
$D^{**}$ control samples, respectively. In each of the four
signal channels, we describe the data as the sum of seven components:
$D\taum\nutb$, $\Dstar\taum\nutb$, $D\ellm\nulb$, $\Dstar\ellm\nulb$, $D^{**}(\ellm/\taum)\nub$,
charge-crossfeed, and combinatorial background. The four $D^{**}$ control samples
are described as the sum of five components: $D^{**}(\ellm/\taum)\nub$, $D(\ellm/\taum)\nub$,
$\Dstar(\ellm/\taum)\nub$, charge-crossfeed, and combinatorial background.
Each of these components is described by one of the two PDFs
given in Section~\ref{sub:pdf}, with the numerical parameters
of the 32 PDFs determined from independent MC samples.
The charge-crossfeed components in the two $D\ellm$
signal channels are described by a single PDF, with common parameters for $\Dz\ellm$ and $\Dp\ellm$, as are the two
$\Dstar\ellm$ charge-crossfeed components and the four $\ds\ellm\piz$
components; the four combinatorial background components in the
signal channels are described by a single PDF with common parameters, as are the four
in the $D^{**}$ control samples.

$B\to\ds\taum\nutb$ events feeding up into the $D^{**}$ control
samples are expected to contribute $1.8\pm 0.6$ events in the
four channels together, so these events are combined with the light
lepton contribution. In both the control samples and in the signal channels,
$B\to D^{**}\taum\nutb$ events are expected to contribute $3.5\%$--$4.5\%$ of
the total $D^{**}$ yield; these events are combined with the
light lepton contribution, and the amount of $D^{**}\taum\nutb$ is
varied as a systematic uncertainty.

\begin{table}
\caption{Components of the signal extraction fit in the signal channels,
and their approximate abundances in our \BB MC sample. The structure of the fit is identical
between the \Bm and \Bzb channels. There are seven components in each of the four
signal channels.}
\label{tab:comps1}
\begin{center}
\begin{tabular}{l l r}\\ \hline\hline
 & & Abundance in \\
Channel          & Source  & \BB MC (\%)\\ \hline
$\Dstarz\ellm$   & $\Dstarz\taum\nutb$ signal           & $5$   \\
                 & $\Dz\taum\nutb$ signal feed-up       & $0.5$ \\
                 & $\Dstarz\ellm\nulb$ normalization       & $90$  \\
                 & $\Dz\ellm\nulb$ feed-up              & $2$   \\
                 & $D^{**}(\ellm/\taum)\nub$ feed-down         & $3$   \\
                 & Charge-crossfeed    & $0.5$ \\
                 & Combinatorial background     & $1$   \\ \hline
$\Dz\ellm$       & $\Dz\taum\nutb$ signal               & $3$   \\
                 & $\Dstarz\taum\nutb$ signal feed-down & $3$   \\
                 & $\Dz\ellm\nulb$ normalization           & $25$  \\
                 & $\Dstarz\ellm\nulb$ feed-down        & $60$  \\
                 & $D^{**}\ellm\nulb$ feed-down         & $2$   \\
                 & Charge-crossfeed    & $2$   \\
                 & Combinatorial background     & $2$   \\ \hline
$\Dstarp\ellm$   & $\Dstarp\taum\nutb$ signal           & $5$   \\
                 & $\Dp\taum\nutb$ signal feed-up       & $0.1$ \\
                 & $\Dstarp\ellm\nulb$ normalization       & $90$  \\
                 & $\Dp\ellm\nulb$ feed-up              & $0.5$   \\
                 & $D^{**}\ellm\nulb$ feed-down         & $3$   \\
                 & Charge-crossfeed    & $0.1$ \\
                 & Combinatorial background     & $2$   \\ \hline
$\Dp\ellm$       & $\Dp\taum\nutb$ signal               & $5$   \\
                 & $\Dstarp\taum\nutb$ signal feed-down & $2$   \\
                 & $\Dp\ellm\nulb$ normalization           & $45$  \\
                 & $\Dstarp\ellm\nulb$ feed-down        & $40$  \\
                 & $D^{**}\ellm\nulb$ feed-down         & $6$   \\
                 & Charge-crossfeed    & $1$   \\
                 & Combinatorial background     & $3$   \\ \hline\hline
\end{tabular}
\end{center}
\end{table}

\begin{table}
\caption{Components of the signal extraction fit in the $D^{**}$
control sample channels, and their approximate
abundances in our \BB MC sample. The structure of the fit is identical
between the \Bm and \Bzb channels. There are five components in each of the four $D^{**}$
control sample channels.}
\label{tab:comps2}
\begin{center}
\begin{tabular}{l l r}\\ \hline\hline
 & & Abundance in \\
Channel          & Source  & \BB MC (\%)\\ \hline
$\Dstarz\piz\ellm$ & $D^{**}\ellm\nulb$                & $60$  \\
              & $\Dstarz\ellm\nulb$ feed-up       & $18$  \\
              & $\Dz\ellm\nulb$ feed-up           & $2$   \\
              & Charge-crossfeed & $1$   \\
              & Combinatorial background  & $20$  \\ \hline
$\Dz\piz\ellm$     & $D^{**}\ellm\nulb$                & $70$  \\
              & $\Dstarz\ellm\nulb$ feed-up       & $10$   \\
              & $\Dz\ellm\nulb$ feed-up           & $3$   \\
              & Charge-crossfeed & $2$   \\
              & Combinatorial background  & $15$  \\ \hline
$\Dstarp\piz\ellm$ & $D^{**}\ellm\nulb$                & $65$  \\
              & $\Dstarp\ellm\nulb$ feed-up       & $20$  \\
              & $\Dp\ellm\nulb$ feed-up           & $0.1$   \\
              & Charge-crossfeed & $0.1$   \\
              & Combinatorial background  & $15$  \\ \hline
$\Dp\piz\ellm$     & $D^{**}\ellm\nulb$                & $60$  \\
              & $\Dstarp\ellm\nulb$ feed-up       & $6$   \\
              & $\Dp\ellm\nulb$ feed-up           & $3$   \\
              & Charge-crossfeed & $1$   \\
              & Combinatorial background  & $30$  \\ \hline\hline
\end{tabular}
\end{center}
\end{table}

The fit has 18 free parameters: four signal branching-fraction
ratios $R$, one for each \ds meson; four $B\to\ds\ellm\nulb$
normalization yields; four $B\to D^{**}\ellm\nulb$
background yields; four combinatorial background yields,
one in each of the four $D^{**}$ control samples; two
parameters describing $\Dstar\Rightarrow D$ feed-down, one
for charged $B$ modes and one for neutral $B$ modes. The
combinatorial background yields in the four signal channels
are fixed in the fit to the expected value from simulation, as are
the charge-crossfeed backgrounds in both the signal channels
and $D^{**}$ control samples; variation of these backgrounds
is treated as a systematic uncertainty below.

We also perform a second, \Bm--\Bzb constrained, fit, by
requiring $R(\Dp)=R(\Dz)$ and
$R(\Dstarp)=R(\Dstarz)$,\footnote{This constraint follows from isospin
symmetry in both the signal and normalization modes but is more general.}
reducing the number of free parameters
to 16.

\subsection{Probability Density Functions}\label{sub:pdf}
We construct an empirical model of the two-dimensional (\mmiss, \pstarl) PDF as the
product of two terms: a one-dimensional
function to describe the \pstarl distribution, discussed
in Section~\ref{sub:oned}; and a \pstarl-dependent ``resolution'' function
to describe the \mmiss distribution, to be discussed
in Section~\ref{sub:twod}.
For processes in which the only missing particle is a single
neutrino, the true \mmiss spectrum is a delta function located at zero and the
observed distribution is a pure resolution function. For components with multiple
missing particles, the observed \mmiss distribution is the convolution of the
physical \mmiss spectrum with our detector resolution. The PDFs
presented below are used to describe both of these physical cases, with different
numerical parameters describing the different behaviors; these two
PDFs are flexible enough to describe the variety of physical and
resolution processes needed in this analysis.

\subsubsection{One-dimensional \pstarl Parameterization}\label{sub:oned}
We use a generalized form of a Gaussian to model the \pstarl distribution.
The Gaussian distribution,

\begin{equation}
\mathcal G(\pstarl)\propto \exp\left (-\frac{1}{2}\cdot\left |\frac{\pstarl-p_0}{\sigma}\right |^2\right )~,
\end{equation}

\noindent has the same general properties as our distributions: it rises smoothly
from zero to a peak value and then falls smoothly back to zero again. Here, $p_0$ represents
the value of $\pstarl$ for which $\mathcal G$ peaks and $\sigma$ represents the width
of the Gaussian distribution.

This gross agreement
is not enough, however, so we define a modified Gaussian function,

\begin{equation}
\mathcal H(\pstarl)\propto \exp\left (-\left |\frac{\pstarl-p_0}{\sigma(\pstarl)}\right |^{\nu(\pstarl)}\right )~,
\end{equation}

\noindent where, for convenience, we have absorbed the constant factor of $2$ into
the definition of $\sigma(\pstarl)$.
By allowing the width and exponent of the Gaussian to be functions of \pstarl, we
are able to describe a greater variety of shapes. Specifically, we take
$\nu(\pstarl)$ to be a linear function,

\begin{equation}
\nu(\pstarl)=\nu_L + \frac{\nu_H-\nu_L}{2.4\gevc}\cdot \pstarl~\,
\end{equation}

\noindent where $\nu_L$ and $\nu_H$ are the values of the exponential term at the low
and high endpoints of \pstarl, fixed at zero and $2.4\gevc$, respectively.
Similarly, we parameterize $\sigma(\pstarl)$ as a bilinear function,

\begin{equation}
\sigma(\pstarl)=
  \begin{cases}
  \sigma_L+\frac{\displaystyle \sigma_0-\sigma_L}{\displaystyle p_0}\cdot \pstarl & \pstarl<p_0 \\
  \sigma_0+\frac{\displaystyle \sigma_H-\sigma_0}{\displaystyle 2.4\gevc -p_0}\cdot (\pstarl -p_0) & \pstarl>p_0
  \end{cases}~,
\end{equation}

\noindent where $\sigma_L$, $\sigma_0$, and $\sigma_H$ represent the widths of the
Gaussian at $\pstarl=0$, $\pstarl=p_0$, and $\pstarl=2.4\gevc$, respectively.
Even though this parameterization is discontinuous at the point
$\pstarl=p_0$, the resulting function $\mathcal H(\pstarl)$ remains
smooth since the numerator in the exponent, $(\pstarl-p_0)$,
goes to zero at the same point.

The \pstarl parameterization therefore has six free parameters: $p_0$, the peak;
$\nu_L$ and $\nu_H$, describing the exponential term; and $\sigma_L$, $\sigma_0$, and
$\sigma_H$, describing the width. When performing fits using this PDF, we
integrate $\mathcal H$ numerically to compute the normalization.

\subsubsection{Two-dimensional PDF Parameterization}\label{sub:twod}
We construct two types of two-dimensional PDF, $\mathcal P_1(\pstarl,\mmiss)$ and
$\mathcal P_2(\pstarl,\mmiss)$ by multiplying the model of the lepton spectrum
above by a ``resolution'' function in \mmiss, where the resolution is a function
of \pstarl. Allowing the parameters of the resolution function to be functions
of \pstarl produces a correlation between the two fit variables, and it is
these parameters which allow the PDFs to describe such a wide variety of shapes.

Using the model of the lepton spectrum $\mathcal H(\pstarl)$ introduced above,
we construct the PDFs as:

\begin{multline}
\mathcal P_1(\pstarl,\mmiss) \equiv \mathcal H(\pstarl)\ \times\\
 \Big [f_1(\pstarl)\mathcal G_1(\pstarl;\mmiss) + \\
 \Big (1-f_1(\pstarl)\Big)\mathcal G_2(\pstarl;\mmiss)\Big]
\end{multline}

\noindent and

\begin{multline}
\mathcal P_2(\pstarl,\mmiss) \equiv \mathcal H(\pstarl)\ \times \\
  \Big[f_1(\pstarl)\mathcal G_1(\pstarl;\mmiss) + \\
    f_2(\pstarl)\mathcal G_b(\pstarl;\mmiss) + \\
    \Big(1-f_1(\pstarl)-f_2(\pstarl)\Big)\mathcal G_2(\pstarl;\mmiss)\Big]\ .
\end{multline}

\noindent Here, the functions $\mathcal G_1$ and $\mathcal G_2$ 
are Gaussians and $\mathcal G_b$ is a bifurcated
Gaussian (Gaussian with different $\sigma$ parameters on either side of
the mean), respectively; all are functions of \mmiss, with parameters dependent on
\pstarl.

The \pstarl dependence of the various parameters of $\mathcal G_{1,2}$ and
$\mathcal G_b$ is listed in Table~\ref{tab:pars}. The total number of free parameters
for $\mathcal P_1$ is 18: six for $\mathcal H(\pstarl)$, five each for $\mathcal G_1$
and $\mathcal G_2$, and two for $f_1$. The total number of free parameters for
$\mathcal P_2$ is 24: six for $\mathcal H(\pstarl)$, five each for $\mathcal G_1$ and
$\mathcal G_2$, four for $\mathcal G_b$, and two each for $f_1$ and $f_2$.

\begin{table}
\caption{\pstarl dependence of the \mmiss PDF parameterizaion. The form of $f_2$ is chosen to
allow the $\mathcal G_b$ term to contribute at low \pstarl, but to drive this term
rapidly to zero as \pstarl increases. The form of $\sigma_H$ is chosen to allow for
a long tail towards high \mmiss at low \pstarl, but to drive this term rapidly to zero
as \pstarl increases (note that there is no problem having $\sigma$ approach zero since
the amplitude of this term goes to zero as well; the result is finite and well-behaved).
$N_\mathrm{par}$ gives the number of free parameters for each term separately.}
\label{tab:pars}
\begin{center}
\begin{tabular}{l l l l}\\ \hline\hline
Function       & Parameter  & Dependence on \pstarl & $N_\mathrm{par}$ \\ \hline
$\mathcal G_{1,2}$ & mean       & quadratic    & 3 \\ $\mathcal G_{1,2}$ & $\sigma$   & linear       & 2 \\ \hline
$\mathcal G_b$ & mean       & constant     & 1 \\
$\mathcal G_b$ & $\sigma_L$ & constant     & 1 \\
$\mathcal G_b$ & $\sigma_H$ & \rule{0pt}{22pt}\rule[-14pt]{0pt}{18pt}$\sigma_{H0}\cdot\left [1-\left (\frac{\displaystyle \pstarl}{\displaystyle 2.4\gevc}\right )^\phi\right ]$ & 2 \\ \hline
$\mathcal P_{1,2}$ & $f_1$      & linear       & 2 \\ \hline
$\mathcal P_2$ & $f_2$      & \rule{0pt}{22pt}\rule[-14pt]{0pt}{18pt}$F\cdot\left (\frac{\displaystyle 2.4\gevc -\pstarl}{\displaystyle 2.2\gevc}\right )^\theta$ & 2 \\ \hline\hline
\end{tabular}
\end{center}
\end{table}

We use the simpler PDF, $\mathcal P_1$, to model
most of the semileptonic fit components (22 out of 32),
as well as the charge-crossfeed and combinatorial backgrounds.
For the remaining
ten components, however, the more complicated parameterization $\mathcal P_2$ is required
to adequately describe the \mmiss tail. Eight of these components are the ones in which the
only missing particle is a single neutrino,

\begin{center}
\begin{tabular}{l @{\hspace{1cm}} l}
$\Dstarz\ellm\nulb\Rightarrow\Dstarz\ellm$ & $\Dz\ellm\nulb\Rightarrow\Dz\ellm$\\
$\Dstarp\ellm\nulb\Rightarrow\Dstarp\ellm$ & $\Dp\ellm\nulb\Rightarrow\Dp\ellm$\\
$D^{**}\ellm\nulb\Rightarrow\Dstarz\piz\ellm$ & $D^{**}\ellm\nulb\Rightarrow\Dz\piz\ellm$\\
$D^{**}\ellm\nulb\Rightarrow\Dstarp\piz\ellm$ & $D^{**}\ellm\nulb\Rightarrow\Dp\piz\ellm$~,\\
\end{tabular}
\end{center}

\noindent and the remaining two are components in which a single neutrino and a soft
$\piz$ or $\gamma$ are missing,

\begin{center}
\begin{tabular}{l @{\hspace{1cm}} l}
$\Dstarz\ellm\nulb\Rightarrow\Dz\ellm$ & $\Dstarp\ellm\nulb\Rightarrow\Dp\ellm$~.\\
\end{tabular}
\end{center}

\begin{figure}
\includegraphics[width=3.1in]{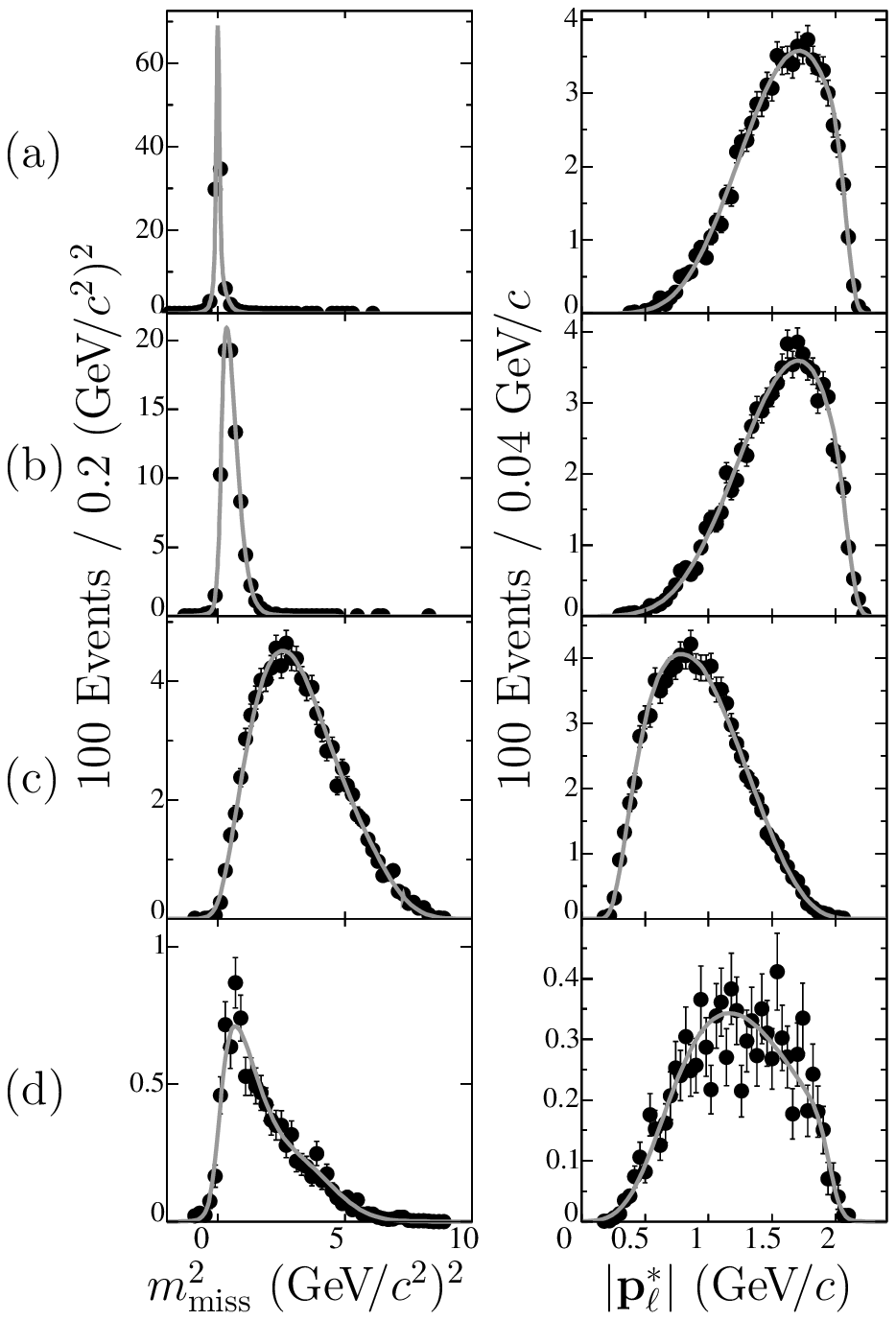}
\caption{Projections of the PDF from fits to MC samples. The left plots show projections
onto \mmiss, while the right plots show projections onto \pstarl. Shown are projections
for four of the PDFs used in the fit: (a) $\Dstarz\ellm\nulb\Rightarrow\Dstarz\ellm$,
(b) $\Dstarz\ellm\nulb\Rightarrow\Dz\ellm$, (c) $\Dstarz\taum\nutb\Rightarrow\Dstarz\ellm$, and
(d) $D^{**}\ellm\nulb\Rightarrow\Dstarz\ellm$. The MC sample is shown as points, and
the projection of the fit is shown as a curve. Note the sharp peak at $\mmiss=0$
in (a), while the peak in (b) is somewhat spread out and shifted to larger values
of \mmiss because of the lost \piz or $\gamma$ from \Dstarz decay.}
\end{figure}

\subsection{Crossfeed Constraints}\label{sub:constraints}
We apply a number of constraints in the fit, relating the event yields
between different reconstruction channels in order to make use of all available
information. These constraints help to maximize our sensitivity, particularly
to the $B\to D\taum\nutb$ signals where the dominant backgrounds are due to
feed-down. There are 20 such constraints in the fit, corresponding to 20
different ways in which a true $B\to D/\Dstar/D^{**}\ellm\nulb$ event can be
reconstructed with the wrong final-state meson, either as feed-down
($\Dstar\Rightarrow D$ and $D^{**}\Rightarrow D/\Dstar$) or as feed-up
($D\Rightarrow\Dstar/\ds\piz$ and $\Dstar\Rightarrow\ds\piz$).

These constraints are implemented in the fit by requiring that
the number of events of type $j$ correctly reconstructed in the
$i^\mathrm{th}$ channel ($N_{ij}$) is related to the number of events
of type $j$ reconstructed in a crossfeed channel $i'$ ($N_{i'j}$)
by

\begin{equation}
N_{i'j}\equiv N_{ij}\cdot f_{i\to i',j}~,
\end{equation}

\noindent where $f_{i\to i',j}$ is a crossfeed constraint relating the two yields.
The crossfeed constraints $f_{i\to i',j}$ are linearly related to the misreconstruction
probability. For feed-down processes, in which the probability to lose a
low-momentum \piz or $\gamma$ is high, $f_{i\to i',j}$ typically takes values
between $0.2$ and $1.0$; for feed-up processes, in which the probability
to reconstruct a fake \piz or $\gamma$ in a narrow mass window is
low, $f_{i\to i',j}$ typically takes values between $0.01$ and $0.1$.

The values for most of the $f_{i\to i',j}$ terms are taken from
simulation, but, in order to reduce systematic effects, the
values of the dominant feed-down components, $B\to\Dstar\ellm\nulb$
reconstructed in the $D\ellm$ signal channels, are left free in the fit to data.
We also use the floating values of these \Dstar feed-down constraints
to apply a small first-order correction to the corresponding signal feed-down
constraints describing $B\to\Dstar\taum\nutb$ reconstructed in the $D\ellm$
channels; in this way, we use the high-statistics $\Dstar\ellm\nulb$
samples to improve our knowledge of the signal feed-down probability.

\subsection{Projections of the Fit to Data}
\begin{figure}
\includegraphics[width=3.1in]{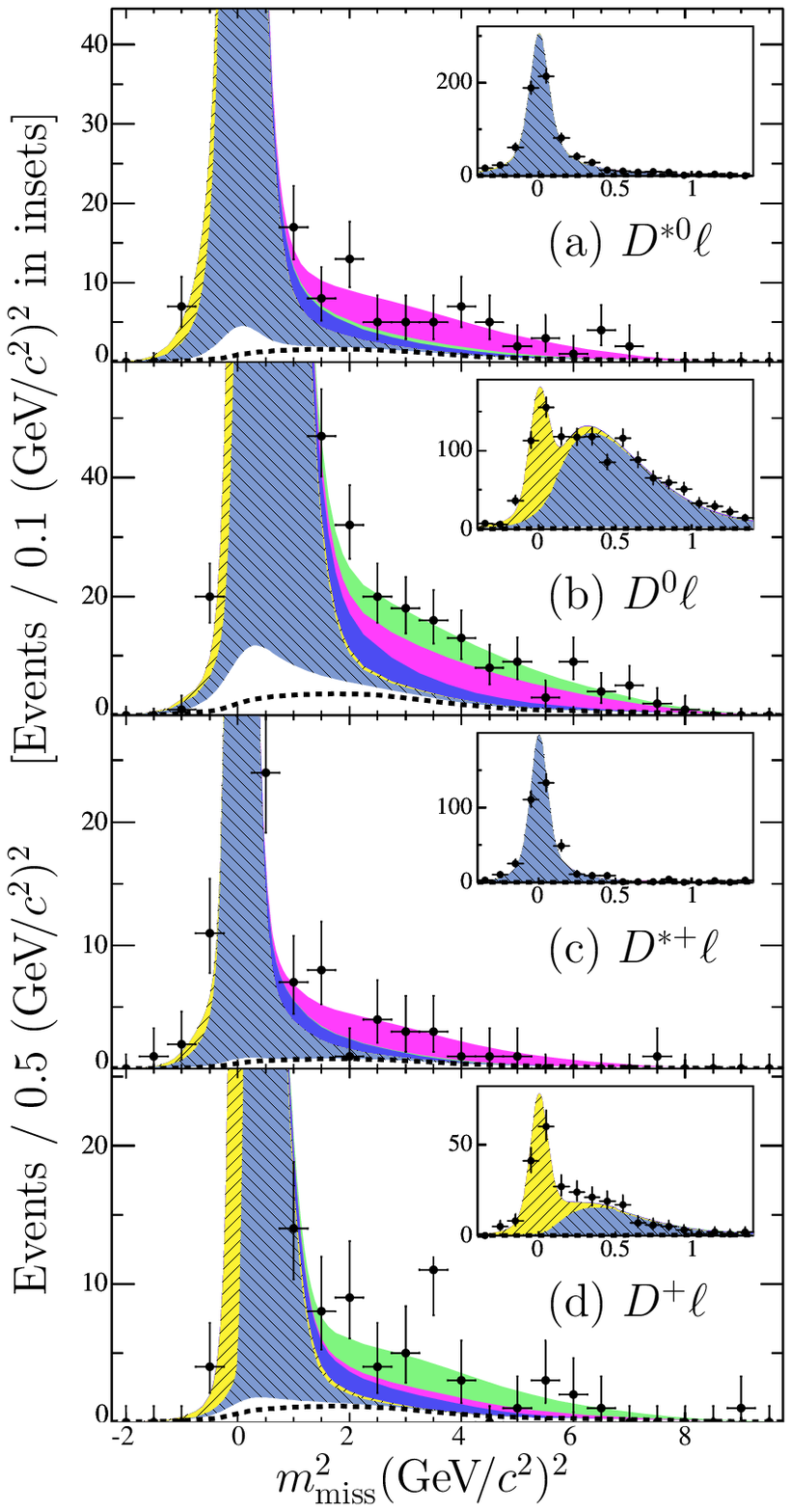}
\caption {(Color online) Distributions of events and fit projections in \mmiss 
for the four final states:
$\Dstarz\ell^-$, $\Dz\ell^-$, $\Dstarp\ell^-$, and $\Dp\ell^-$. 
The normalization region $\mmiss\approx 0$ is shown with finer binning in the insets. 
The fit components are combinatorial background (white, below dashed line), 
charge-crossfeed background (white, above dashed line), the $B\to D\ellm\nulb$ 
normalization mode (// hatching, yellow), the $B\to\Dstar\ellm\nulb$ normalization
mode ($\backslash\backslash$ hatching, light blue), $B\to D^{**}\ellm\nulb$ background (dark, or blue),
the $B\to D\taum\nutb$ signal (light grey, green), and the $B\to\Dstar\taum\nutb$ signal
(medium grey, magenta). The fit shown incorporates the \Bm--\Bzb constraints.
}\label{fig:fit}
\end{figure}

\begin{figure}
\includegraphics[width=3.1in]{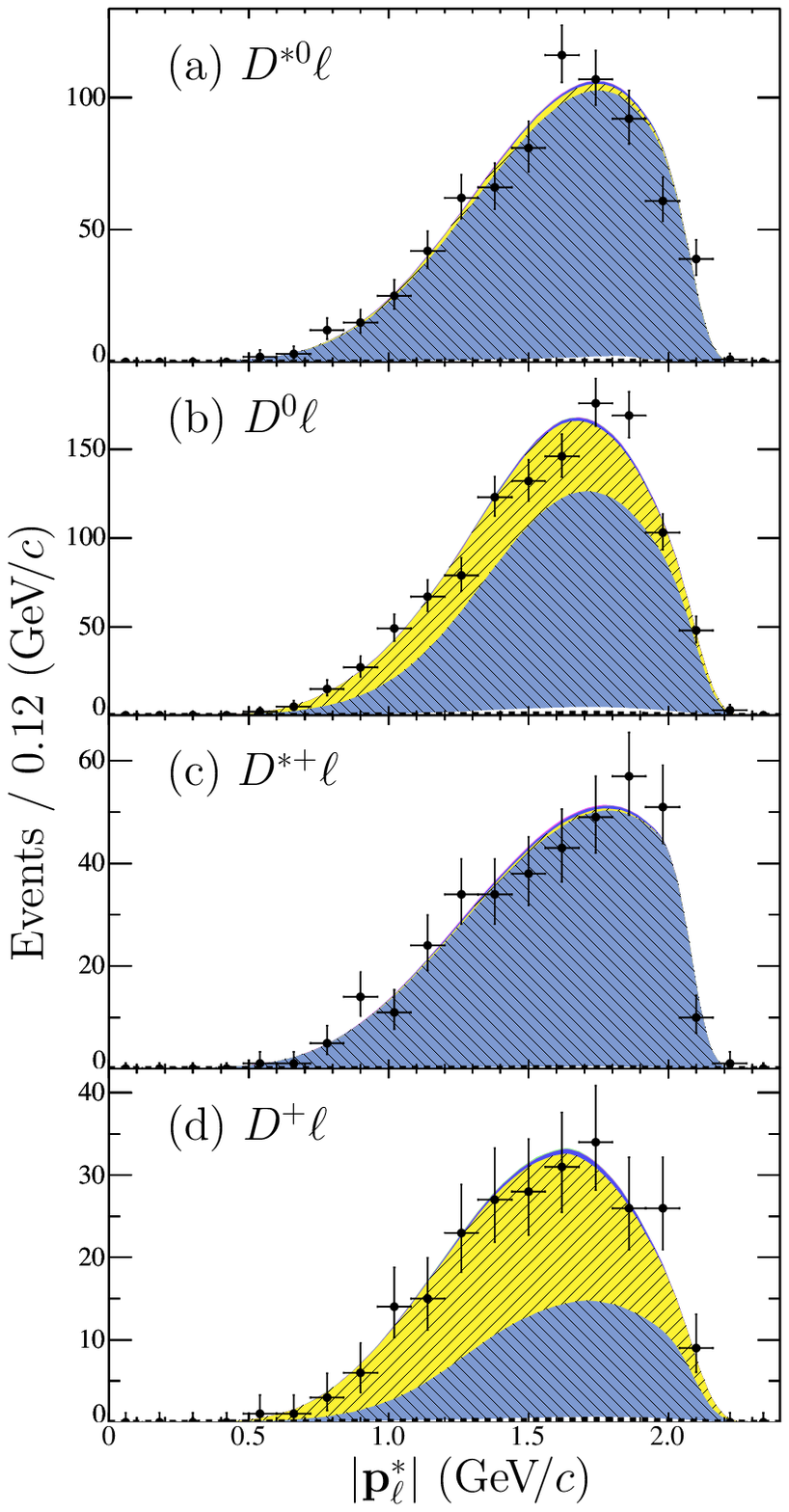}
\caption{(Color online) Distributions of events and fit projections
in \pstarl for the four final states
$\Dstarz\ell^-$, $\Dz\ell^-$, $\Dstarp\ell^-$, and $\Dp\ell^-$,
shown in the normalization region, $\mmiss<1\ (\gevccnosp)^2$.
The fit components are shaded as in Fig.~\ref{fig:fit}.}
\label{fig:pstarnorm}
\end{figure}

\begin{figure}
\includegraphics[width=3.1in]{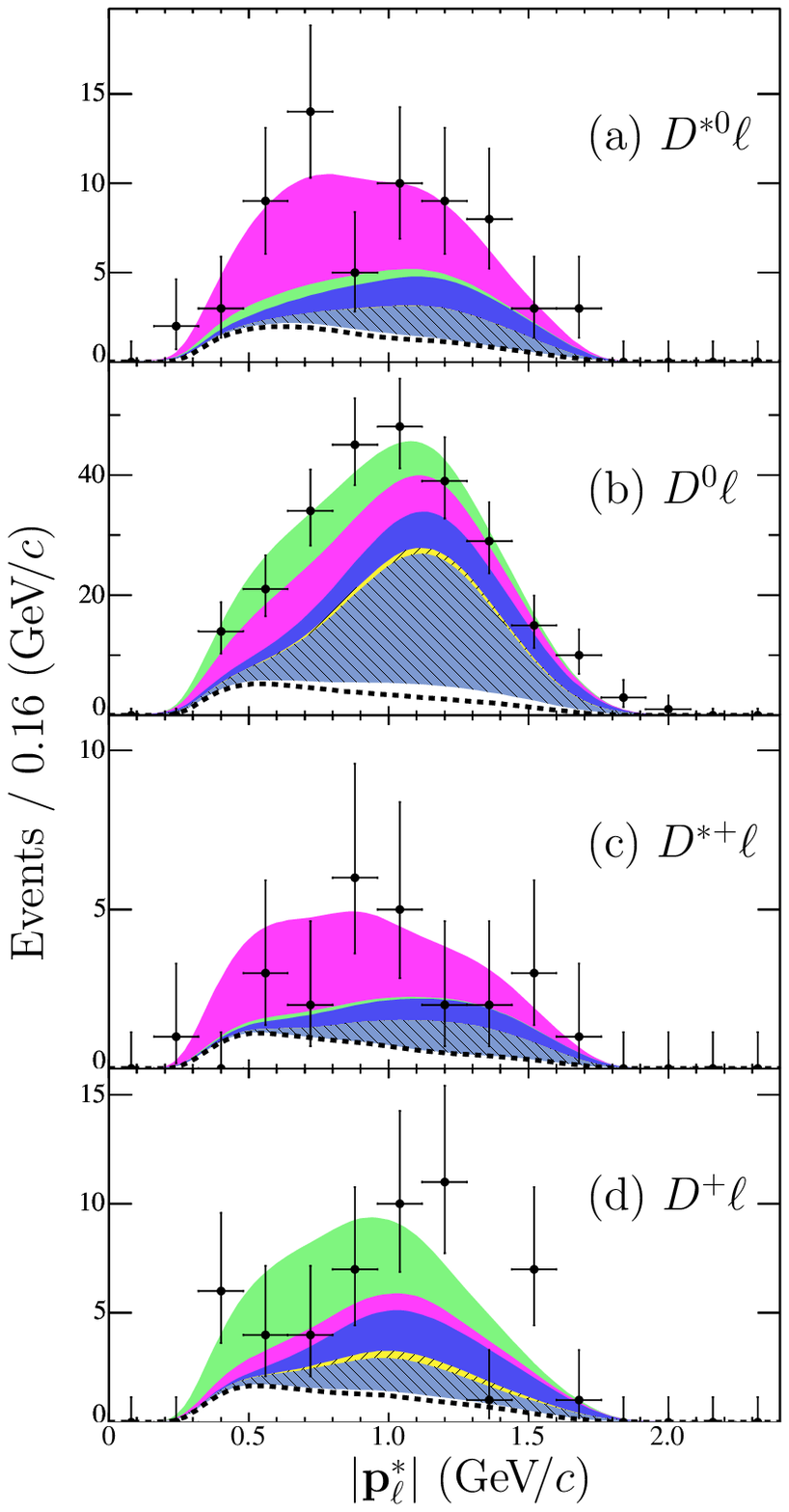}
\caption{(Color online) Distributions of events and fit projections
in \pstarl for the four final states
$\Dstarz\ell^-$, $\Dz\ell^-$, $\Dstarp\ell^-$, and $\Dp\ell^-$,
shown in the signal region, $\mmiss>1\ (\gevccnosp)^2$.
The fit components are shaded as in Fig.~\ref{fig:fit}.}
\label{fig:pstarsig}
\end{figure}

\begin{figure}
\includegraphics[width=3.1in]{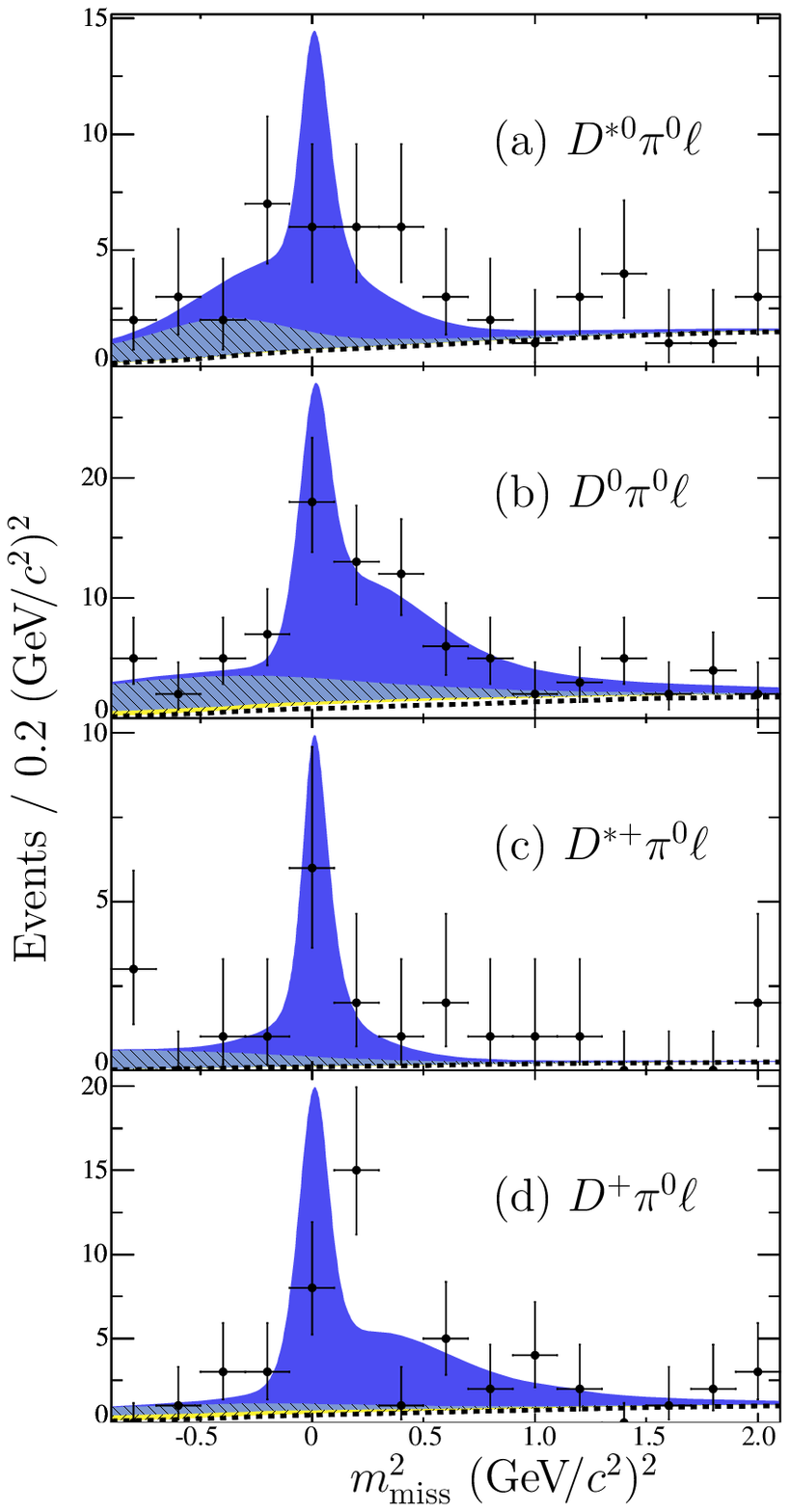}
\caption{(Color online) Distributions of events and fit projections
in \mmiss for the four $D^{**}$ control samples
$\Dstarz\piz\ell^-$, $\Dz\piz\ell^-$, $\Dstarp\piz\ell^-$, and $\Dp\piz\ell^-$.
The fit components are shaded as in Fig.~\ref{fig:fit}.}
\label{fig:mm2dss}
\end{figure}

\begin{figure}
\includegraphics[width=3.1in]{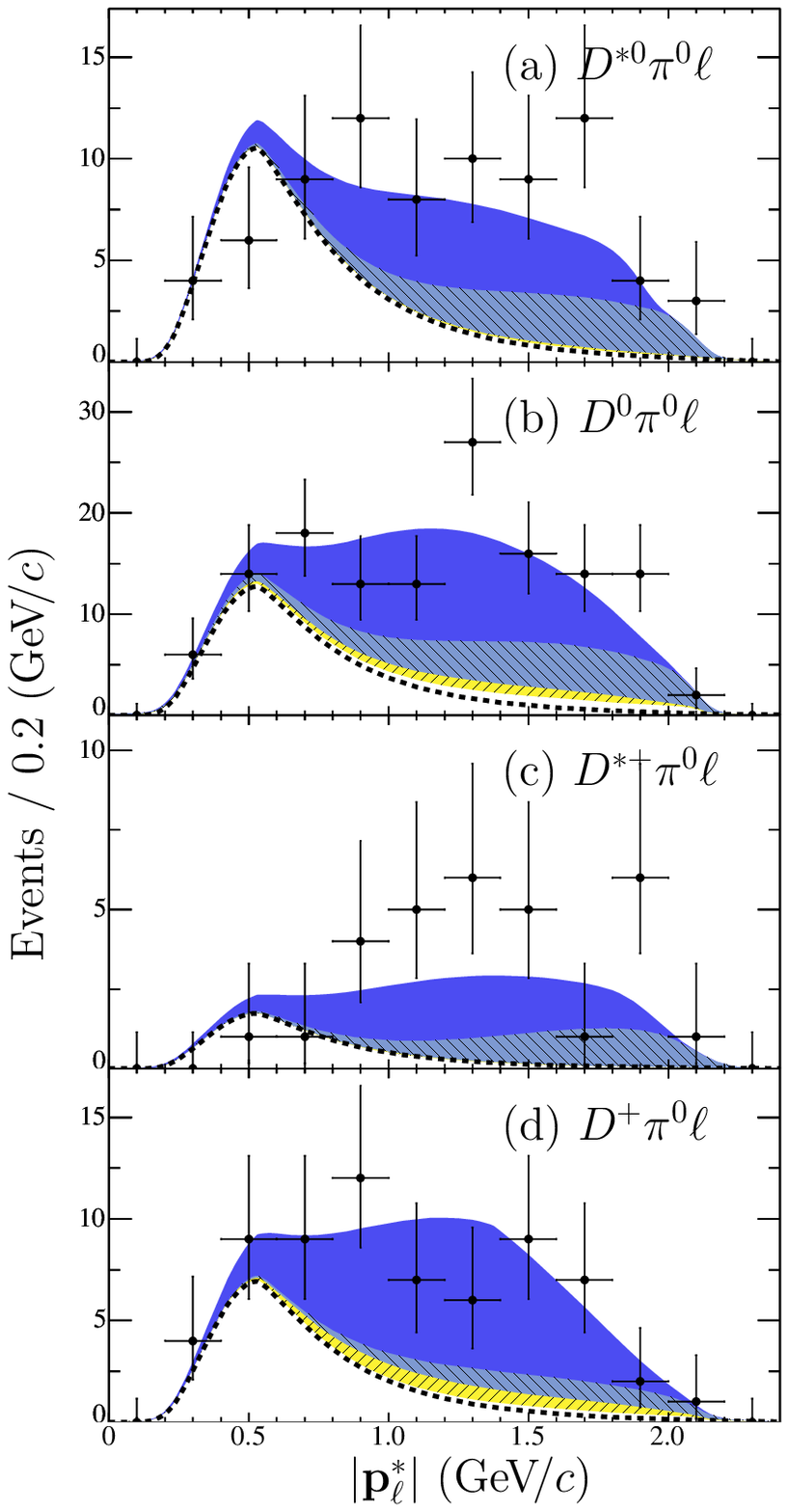}
\caption{(Color online) Distributions of events and fit projections
in \pstarl for the four $D^{**}$ control samples
$\Dstarz\piz\ell^-$, $\Dz\piz\ell^-$, $\Dstarp\piz\ell^-$, and $\Dp\piz\ell^-$.
The fit components are shaded as in Fig.~\ref{fig:fit}.}
\label{fig:pstardss}
\end{figure}

Figures~\ref{fig:fit}--\ref{fig:pstardss} show projections of
the \Bm--\Bzb-constrained fit. Figure~\ref{fig:fit} shows projections
in \mmiss for the four signal channels, showing both the low \mmiss
region, which is dominated by the normalization modes $B\to\ds\ellm\nulb$,
and the high \mmiss region, which is dominated by the signal modes
$B\to\ds\taum\nutb$. Figures~\ref{fig:pstarnorm} and \ref{fig:pstarsig}
show projections in \pstarl for the normalization and signal regions,
respectively, and Figs.~\ref{fig:mm2dss} and \ref{fig:pstardss}
show projections of both \mmiss and \pstarl for the four
$D^{**}$ control samples. In all cases, we see that the fit
does a reasonable job of describing the observed event sample,
both in background-dominated and signal-dominated regions.

\section{Signal Extraction and Normalization}
The fit described in Section~\ref{sec:fit} directly measures, for each
signal mode, the ratio of the number of signal events in the data sample,
$N_\mathrm{sig}$, to the number of corresponding normalization events,
$N_\mathrm{norm}$. We measure the signal branching-fraction ratios $R$ as

\begin{equation}
R\equiv\frac{N_\mathrm{sig}}{N_\mathrm{norm}}\cdot\frac{1}{\varepsilon_\mathrm{sig}/\varepsilon_\mathrm{norm}}\cdot\frac{1}{\BR(\taum\to\ellm\nulb\nut)}~,
\label{eq:norm}\end{equation}

\noindent where the relative efficiency
$\varepsilon_\mathrm{sig}/\varepsilon_\mathrm{norm}$ is calculated 
from  signal MC samples as

\begin{equation}
\varepsilon_\mathrm{sig}/\varepsilon_\mathrm{norm}\equiv
\frac{N^\mathrm{reco}_\mathrm{sig}/N^\mathrm{gen}_\mathrm{sig}}
     {N^\mathrm{reco}_\mathrm{norm}/N^\mathrm{gen}_\mathrm{norm}}~.
\end{equation}

\noindent Here, the $N^\mathrm{gen}$ are the numbers of simulated
events, and the $N^\mathrm{reco}$ are the numbers of reconstructed
events, including both correctly reconstructed events and contributions
from feed-up or feed-down. Crossfeed is not a large effect, however,
because both the numerator and denominator in this relative efficiency
receive crossfeed contributions, and the net result tends to cancel (this
cancellation is not exact, since the \Dstar momentum spectra are not
identical between signal and normalization modes, but these differences
are already accounted for in our normalization procedure).

Signal efficiencies are given in Table~\ref{tab:eff}.
The relative efficiencies for the two $B\to D\taum\nutb$ modes
are much larger than unity because of the $q^2$ cut, which
is $\approx 98\%$ efficient for signal events but rejects about
$50\%$ of the $B\to D\ellm\nulb$ normalization events, as seen in
Fig.~\ref{fig:cln_q2}(a). The $q^2$ cut has a similar, but
less pronounced, effect on the \Dstar modes, but, due to the lower
efficiency for identifying secondary leptons in the signal modes,
the resulting relative efficiency is close to unity.

\begin{table}
\caption{Relative signal efficiencies
$\varepsilon_\mathrm{sig}/\varepsilon_\mathrm{norm}$ for the four signal
modes.}
\label{tab:eff}
\begin{center}
\begin{tabular}{l l}\\ \hline\hline
Signal mode & $\varepsilon_\mathrm{sig}/\varepsilon_\mathrm{norm}$ \\ \hline
$\Bm\to\Dz\taum\nutb$ & $1.85\pm 0.02$\\
$\Bm\to\Dstarz\taum\nutb$ & $0.99\pm 0.01$\\
$\Bzb\to\Dp\taum\nutb$ & $1.83\pm 0.03$\\
$\Bzb\to\Dstarp\taum\nutb$ & $0.91\pm 0.01$\\ \hline\hline
\end{tabular}
\end{center}
\end{table}

\section{Systematic Uncertainties}
Table~\ref{tab:syssum} summarizes all of the systematic
uncertainties considered in this analysis. Because our signal
is extracted and normalized relative to $B\to\ds\ellm\nulb$,
many sources of systematic uncertainty---especially those
related to reconstruction efficiency---are expected to cancel,
either partially or completely, when we take the ratio.

We describe the individual contributions to the systematic
uncertainty below. We divide the systematics into two broad
categories: additive and multiplicative. Additive systematic
uncertainties are those which affect the fit yields and
therefore reduce the significance of the measured signals.
Multiplicative uncertainties affect the normalization of
the signals and the numerical results but not
the significance.

\begin{table*}
\caption{Contributions to the total systematic uncertainty. The additive systematic
uncertainties represent uncertainties on the fit yield, and therefore reduce
the statistical significance of the results. The multiplicative systematic uncertainties 
represent uncertainties on the normalization, so they affect the
numerical results but not the statistical significance. The first four
columns summarize errors on the individual branching-fraction ratios; the last
two columns summarize errors on the \Bm--\Bzb constrained measurement. The
totals here refer to errors on the branching-fraction ratios $R$; the errors
on $\BR(B\to\ds\ellm\nulb)$ (discussed in Section~\ref{results}) only apply to the absolute branching
fractions, and are not included in the quoted total error.}
\label{tab:syssum}
\begin{center}
\begin{tabular*}{2.0\columnwidth}{@{\extracolsep{\fill}}l l l l l l l}\\ \hline\hline
Source & \multicolumn{6}{c}{Fractional uncertainty (\%)} \\
 & $\Dz\tau\nu$ & $\Dstarz\tau\nu$ & $\Dp\tau\nu$ & $\Dstarp\tau\nu$ & $D\tau\nu$ & $\Dstar\tau\nu$\\ \hline
& \multicolumn{6}{c}{Additive systematic uncertainties}\\
MC stat. (PDF shape)                  &11.5 & 8.4 & 4.5 & 1.8 & 6.9 & 4.7\\
MC stat. (constraints)                & 4.2 & 1.9 & 6.1 & 1.3 & 3.6 & 1.4\\
Comb. BG modeling                     & 7.5 & 4.1 &11.5 & 2.6 & 9.1 & 2.9\\
$D^{**}$ modeling                     & 5.7 & 0.5 & 1.6 & 0.2 & 3.0 & 0.4\\
$B\to\Dstar$ form factors             & 1.9 & 0.7 & 0.8 & 0.2 & 1.4 & 0.4\\
$B\to D$ form factors                 & 0.2 & 0.7 & 0.6 & 0.2 & 0.3 & 0.4\\
\mmiss tail modeling                  & 1.5 & 0.5 & 1.2 & 0.4 & 1.6 & 0.1\\
\piz crossfeed constraints            & 0.5 & 1.1 & 0.5 & 0.9 & 0.5 & 1.0\\
$D^{**}$ feed-down                    & 0.4 & 0.1 & 0.1 & 0.3 & 0.2 & 0.2\\
$D^{**}\taum\nutb$ abundance          & 0.4 & 1.3 & 0.3 & 0.2 & 0.3 & 0.8\\ \hline
Total additive             & 15.6 &  9.7 & 14.0 &  3.6 & 12.5 &  5.8\\ \hline
& \multicolumn{6}{c}{Multiplicative systematic uncertainties}\\
MC stat. (efficiency)      & 1.2 & 1.1 & 1.5 & 1.1 & 1.0 & 0.8\\
Bremsstrahlung/FSR         & 0.6 & 0.5 & 0.3 & 0.4 & 0.4 & 0.5\\
Tracking $\varepsilon$     & 0.0 & 0.0 & 0.0 & 0.0 & 0.0 & 0.0\\
$e$ PID $\varepsilon$      & 0.5 & 0.5 & 0.6 & 0.6 & 0.6 & 0.6\\
$\mu$ PID $\varepsilon$    & 0.5 & 0.6 & 0.7 & 0.6 & 0.6 & 0.6\\
$K$ PID $\varepsilon$      & 0.2 & 0.1 & 0.2 & 0.0 & 0.2 & 0.0\\
$\pi$ PID $\varepsilon$    & 0.1 & 0.1 & 0.2 & 0.0 & 0.1 & 0.1\\
\KS\ $\varepsilon$         & 0.1 & 0.0 & 0.1 & 0.1 & 0.1 & 0.0\\
Neutral (\piz and $\gamma$) $\varepsilon$ & 0.0 & 0.0 & 0.0 & 0.1 & 0.0 & 0.0\\
Daughter $\BR$'s    & 0.1 & 0.3 & 0.0 & 0.1 & 0.1 & 0.3\\
$\BR(\taum\to\ellm\nulb\nut)$ & 0.2 & 0.2 & 0.2 & 0.2 & 0.2 & 0.2\\ \hline
Total multiplicative       & 1.6 & 1.5 & 1.8 & 1.4 & 1.4 & 1.3\\ \hline
Total                      & 15.6 &  9.9 & 14.0 &  3.9 & 12.5 &  6.0\\ \hline
$\BR(B\to\ds\ellm\nulb)$ & 10.2 & 7.7 & 9.4 & 3.7 & 6.8 & 3.4\\ \hline\hline
\end{tabular*}
\end{center}
\end{table*}

\subsection{Additive Systematic Uncertainties}
In order to estimate additive systematic uncertainties,
we perform an ensemble of fits to MC event samples. For each source
of uncertainty, we perform a number of tests where we
modify, as appropriate, the fit shapes, crossfeed constraints, and
the combinatorial background yields (all of which are fixed to
MC-derived values in the nominal fit) and perform a signal fit.
By doing a large number
of such tests and studying the distribution of fit results
in these ensembles, we are able to estimate the systematic
uncertainties. In all of these ensembles, we
take the RMS of the observed distribution,
relative to the corresponding mean fit value, as the systematic
uncertainty.

\subsubsection{Monte Carlo Statistics}
In order to study the systematic uncertainties due to limited
Monte Carlo statistics, we perform two ensembles of fits.
In the first ensemble, we perform a variation of the PDF shapes.
Each of the 37 PDFs are independently varied by generating
new values for each of the 18 or 24 shape parameters according
to the uncertainties in the PDF fit, taking into account correlations
between the fitted parameters. In the second ensemble, we vary each
of the feed-up and feed-down constraints according to their
statistical uncertainties.

\begin{figure}
\includegraphics[width=3.1in]{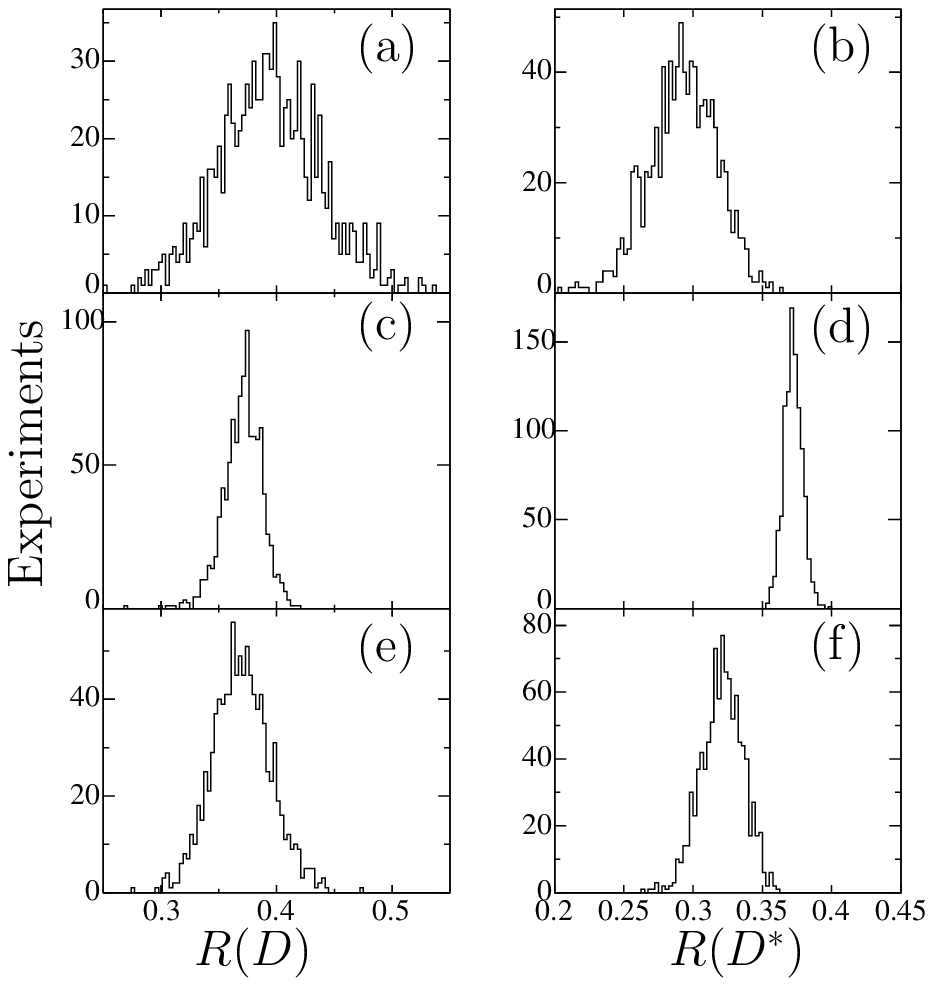}
\caption
{Distributions of fit results for systematic uncertainties due to
Monte Carlo statistics, shown for (a) $\Bm\to\Dz\taum\nutb$,
(b) $\Bm\to\Dstarz\taum\nutb$, (c) $\Bzb\to\Dp\taum\nutb$,
(d) $\Bzb\to\Dstarp\taum\nutb$, (e) $B\to D\taum\nutb$, and
(f) $B\to\Dstar\taum\nutb$. In all
figures, the branching-fraction ratio $R$ is shown.}
\label{fig:sysmcstat}
\end{figure}

Figure~\ref{fig:sysmcstat} shows distributions of fit results
for the ensemble of PDF shape fits.

\subsubsection{Combinatorial Background Modeling}
Table~\ref{tab:combabund} summarizes the physical sources of
combinatorial background considered in this analysis, including their
approximate abundances in our \BB MC sample after all event selection.
In order to study systematic effects,
we perform an ensemble of fits, reweighting events from the
various combinatorial sources.

\begin{table*}
\caption{Sources and approximate abundances of combinatorial background
in our \BB MC sample. All four signal channels are combined here. The third and fourth columns
show what fraction of the $B$ decays in each group have previously been
observed. The fourth column is the product of the second and third, and
indicates how much of the estimated combinatorial background is known from other
measurements.}
\label{tab:combabund}
\begin{center}
\begin{tabular}{l l r r r}\\ \hline\hline
\multicolumn{2}{l}{Source} & \% of total BG & \multicolumn{2}{c}{\% \BR\ measured} \\
\multicolumn{3}{l}{} & (relative) & (absolute) \\ \hline
\multicolumn{2}{l}{$B\to D_s^{(*)+}\ds$ ( + light hadrons)} &  \\
 & \ldots with $\Ds\to\tau\nu$ & 30 & 90 & 27\\
 & \ldots with $\Ds\to\ell\nu (\phi/\eta/\eta')$ & 10 & 90 & 9\\
\multicolumn{2}{l}{$B\to\ds\ds$ ( + light hadrons)} & 35 & 65 & 25\\
\multicolumn{2}{l}{Both $B\to\ds\ell\nu$} & 15 & 100 & 15\\
\multicolumn{2}{l}{Fake lepton} & 5 & 0 & 0\\
\multicolumn{2}{l}{Other misreconstructed} & 5 & 0 & 0\\ \hline
\multicolumn{2}{l}{Total} & & & 75 \\ \hline\hline
\end{tabular}
\end{center}
\end{table*}

In total, the two-body $B$ decays $B\to D_s^{(*)+}D^{(*)(*)}$ and
$B\to\ds\ds$ constitute approximately 45\% of the total
combinatorial background yield, while the three-body decays
$B\to\ds\ds K$ contribute another 15\%.
Branching fractions of most of the relevant two-body $B$ decays (and
some of the three-body decays as well) have previously been
measured. These branching fractions are listed in Table~\ref{tab:combbf},
along with relevant branching fractions of the \Ds\ meson.

\begin{table}
\caption{Branching fractions of $\Ds$ and two- and three-body $B$ decays
contributing to combinatorial background. Measurements are taken
from~\cite{oldPDG}, except ($\dagger$) which are taken from~\cite{babarDD}.
The last column gives the branching fraction used to generate the
\babar\ MC sample, where each number is shown in the same scale
as the corresponding number in the second column.}
\label{tab:combbf}
\begin{center}
\begin{tabular}{l @{ $\to$ } l l @{\ \ \ $($} r @{ $\pm$ } l @ { $)\ \times$ } l r} \hline\hline
\multicolumn{3}{l}{Mode} & \multicolumn{3}{c}{\BR} & MC \\ \hline
$\Ds$&$\tau\nu$ && $6.4$ & $1.5$ & $10^{-2}$ & 7.0\\
$\Ds$&$\eta\ell\nu$ && $2.5$ & $0.7$ & $10^{-2}$ & 2.6\\
$\Ds$&$\eta'\ell\nu$ && $8.9$ & $3.3$ & $10^{-3}$ & 8.9\\
$\Ds$&$\phi\ell\nu$ && $2.0$ & $0.5$ & $10^{-2}$ & 2.0\\
$\Ds$&$\mu\nu$ && $5.0$ & $1.9$ & $10^{-3}$ & 4.6\\ \hline
$\Bp$&$\Dzb\Ds$ && $1.3$ & $0.4$ & $10^{-2}$ & 1.06\\
$\Bp$&$\Dzb\Dss$ && $9$ & $4$ & $10^{-3}$ & 9.1\\
$\Bp$&$\Dstarzb\Ds$ && $1.2$ & $0.5$ & $10^{-2}$ & 1.02\\
$\Bp$&$\Dstarzb\Dss$ && $2.7$ & $1.0$ & $10^{-2}$ & 2.28\\
$\Bp$&$ D^{**0}D_s^{(*)+}$ && $2.7$ & $1.2$ & $10^{-2}$ & 3.0\\
$\Bp$&$\Dzb\Dstarp\Kz$ && $5.2$ & $1.2$ & $10^{-3}$ & 5.2\\
$\Bp$&$\Dstarzb\Dstarp\Kz$ && $7.8$ & $2.6$ & $10^{-3}$ & 7.8\\
$\Bp$&$\Dz\Dzb\Kp$ && $1.37$ & $0.32$ & $10^{-3}$ & 1.9\\
$\Bp$&$\Dstarz\Dstarzb\Kp$ && $5.3$ & $1.6$ & $10^{-3}$ & 5.3\\
$\Bp$&$\Dstarz\Dzb\Kp$ && $4.7$ & $1.0$ & $10^{-3}$ & 4.8\\
$\Bp$&$\Dstarm\Dp\Kp$ && $1.5$ & $0.4$ & $10^{-3}$ & 0.5\\ \hline
$\Bz$&$\Dm\Ds$ && $8.0$ & $3.0$ & $10^{-3}$ & 7.4\\
$\Bz$&$\Dstarm\Ds$ && $1.07$ & $0.29$ & $10^{-2}$ & 1.03\\
$\Bz$&$\Dm\Dss$ && $1.0$ & $0.5$ & $10^{-2}$ & 0.74\\
$\Bz$&$\Dstarm\Dss$ && $1.9$ & $0.5$ & $10^{-2}$ & 1.97\\
$\Bz$&$\Dm\Dz\Kp$ && $1.7$ & $0.4$ & $10^{-3}$ & 1.7\\
$\Bz$&$\Dm\Dstarz\Kp$ && $3.1$ & $0.6$ & $10^{-3}$ & 3.1\\
$\Bz$&$\Dstarm\Dstarz\Kp$ && $1.18$ & $0.20$ & $10^{-2}$ & 1.18\\
$\Bz$&$\Dstarm\Dp\Kz$ && $6.5$ & $1.6$ & $10^{-3}$ & 8.1\\
$\Bz$&$\Dstarm\Dstarp\Kz$ && $8.8$ & $1.9$ & $10^{-3}$ & 8.8\\
$\Bz$&$\Dstarp\Dstarm$ & ($\dagger$) & $8.1$ & $1.2$ & $10^{-4}$ & 8.3\\
$\Bz$&$\Dp\Dstarm$ & ($\dagger$) & $10.4$ & $2.0$ & $10^{-4}$ & 6.7\\
$\Bz$&$\Dp\Dm$ & ($\dagger$) & $2.8$ & $0.7$ & $10^{-4}$ & 2.7\\ \hline\hline
\end{tabular}
\end{center}
\end{table}

To study systematic uncertainties related to combinatorial
background modeling, we perform an ensemble of fits. In each fit, we
reweight events in the simulation. For modes listed in 
Table~\ref{tab:combbf}, we reweight the branching fraction, generating
random weights from a Gaussian distribution based on
the measured value (for decays involving a \Ds\ meson, the weight
is the product of weights for both the $B$ and \Ds\ decays).
For charge-crossfeed events (true $B\to\ds\ellm\nulb$
events where the \btag and signal \ds swap a charged particle), the
dominant systematic uncertainty is not the branching fraction, but
rather the efficiency to reconstruct the \btag with the wrong charge.
We estimate a 10\% uncertainty on the modeling of this process, {\em i.e.},
we generate weights for these events using a Gaussian with a mean
of $1$ and a width of $0.1$.
For double-semileptonic events,
with both $B$ mesons decaying to $\ds\ellm\nulb$, again, the dominant
uncertainty comes from the probability to misreconstruct a \btag
candidate in this event, and we assume a 10\% uncertainty
on this number as well. For events in which the signal lepton is misidentified,
we assign a 10\% uncertainty; the typical fake rate measured in data is
2\%--3\%, with data-simulation discrepancies generally 10\% or less in the
momentum ranges of interest. For all remaining sources of combinatorial
background, including high-multiplicity $B\to Dhh(h\ldots)$ and
$B\to DDhh(h\ldots)$ (where $D$ here represents any charm meson and
$h$ any light meson) and other misreconstructed events,
we assume a 50\% uncertainty in the relevant
rates.

\begin{figure}
\includegraphics[width=3.1in]{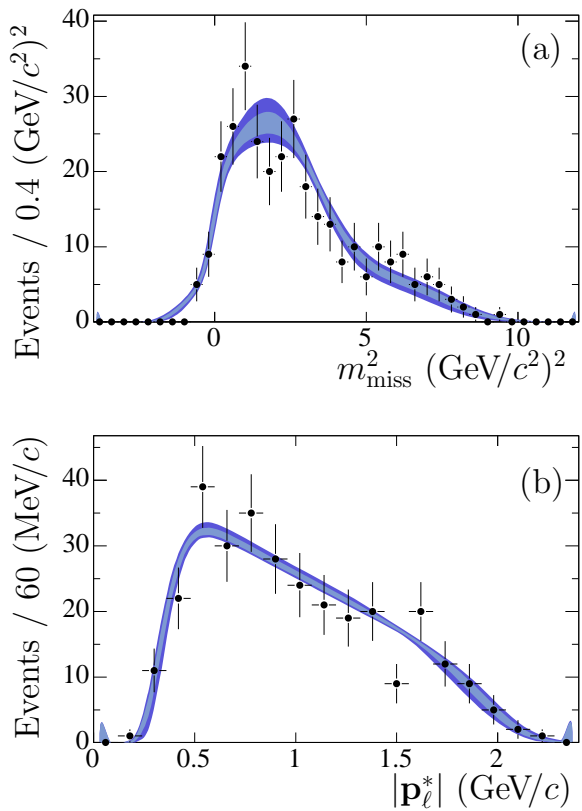}
\caption{Combinatorial background modeling variation for the four signal
channels, showing the projections onto (a) \mmiss, and (b) \pstarl.
In both figures, the MC sample without reweighting is shown as
data points, while the light and dark shaded regions show the
$1\sigma$ and $2\sigma$ envelopes of the ensemble of reweighted
PDF shapes.}
\label{fig:combbands}
\end{figure}

In each test, we fit the reweighted MC sample to generate
new PDF shapes and recalculate the expected yield of combinatorial
events in each channel. Figure~\ref{fig:combbands}
shows the effect of this reweighting on the combinatorial BG PDF
in the signal channels.
We note that the reweighting affects the normalization of the
charge-crossfeed backgrounds but not the shape.

\subsubsection{$D^{**}$ Modeling}\label{sub:dss}
We generate an ensemble of
$D^{**}$ models, sampling from the distribution in Table~\ref{tab:dssbf2}.
This model is based on the current world averages~\cite{hfag,PDG}
but includes information from selected recent measurements~\cite{newdssmodel}
and imposes isospin symmetry between charged and neutral $B$ modes.

For each test, we generate random numbers for the six exclusive
modes ($D$, \Dstar, and the resonant $D^{**}$ states), independently
for \Bp\ and \Bz\ decays. We then saturate the remaining inclusive
$b\to c\ellm\nulb$ rate with the four nonresonant states,
maintaining the Monte Carlo ratio of $0.1:0.3:0.2:0.6$.
Even though we are only interested in the $B\to D^{**}\ellm\nulb$
states, we need to generate distributions of the $B\to\ds\ellm\nulb$
branching fractions to allow for sufficient variations in the
nonresonant states which are used to saturate the total rate.

\begin{table}
\caption{$B\to X_c\ellm\nulb$ branching fractions used in the $D^{**}$
modeling systematic study. The first line, $c\ell\nu$, represents
the inclusive semileptonic branching fraction. For the six lines representing
the $D$, \Dstar, and $D^{**}$ resonant states, the distribution of these
branching fractions is taken to be Gaussian with the given mean and width.
For the last four lines, representing the nonresonant $D^{**}$ states,
the ranges of variation are not shown in this table; their distribution
is determined by the inclusive rate and the other exclusive modes, as
described in the text. The generated branching fractions, $\BR_\mathrm{gen}$,
are the same for charged and neutral $B$ mesons. All numbers are in \%.}
\label{tab:dssbf2}
\begin{center}
\begin{tabular*}{\columnwidth}{@{\extracolsep{\fill}}l l l l l l}\\ \hline\hline
Mode\rule{0pt}{10pt} & $\BR_\mathrm{gen}$ & \multicolumn{2}{c}{\BR (\Bzb)}& \multicolumn{2}{c}{\BR (\Bm)}\\
&  & $\mu$ & $\sigma$& $\mu$ & $\sigma$ \\ \hline
$c\ell\nu$ & 10.4 & 10.17 & 0 & 10.9 & 0\\
$D$ & 2.10 & 2.14 & 0.14 & 2.29 & 0.16 \\
$\Dstar$ & 5.6 & 5.54 & 0.25 & 5.94 & 0.24 \\
$D_1$ & 0.56 & 0.47 & 0.08 & 0.58 & 0.06 \\
$D^*_2$ & 0.37 & 0.35 & 0.07 & 0.46 & 0.08 \\
$D^*_0$ & 0.20 & 0.46 & 0.09 & 0.45 & 0.09 \\
$D_1'$ & 0.37 & 0.85 & 0.20 & 0.83 & 0.20 \\ \hline
$\Dstar\piz$\rule{0pt}{10pt} & 0.1 & 0.03 & --- & 0.029 & ---\\
$D\piz$ & 0.3 & 0.09 & --- & 0.088 & ---\\
$\Dstar\pipm$ & 0.2 & 0.06 & --- & 0.058 & ---\\
$D\pipm$ & 0.6 & 0.18 & --- & 0.175 & ---\\ \hline\hline
\end{tabular*}
\end{center}
\end{table}

For each test, we reweight both the $D^{**}\ellm\nulb$ PDFs
and crossfeed constraints to estimate the systematic uncertainty.

\subsubsection{$B\to\ds$ Form Factors}
We reweight the form factors of both signal $B\to\ds\taum\nutb$
and normalization $B\to\ds\ellm\nulb$ decays. In both cases, we
use the form factor parameterization of Caprini, Lellouch, and
Neubert~\cite{CLN}, with numerical parameters given in
Section~\ref{sec:models}. We reweight signal and normalization
modes simultaneously and generate new PDFs, crossfeed
constraints, and relative efficiencies.

\subsubsection{\mmiss Tail Modeling}
Studies in the two kinematic control samples show acceptable overall
agreement between data and simulation for the \mmiss
resolution [see Fig.~\ref{fig:control}(d)], but suggest that the
simulation may underestimate the ratio of the number of events in
the large \mmiss tail region to the number of events near $\mmiss=0$.
We estimate that this tail component of the resolution may
be underestimated by up to $10\%$.
We study systematic effects related to this by
reweighting events at large \mmiss, greater than
$1\ (\gevccnosp)^2$, up by $10\%$,
modifying the PDF shapes for $B\to D\ellm\nulb$ and
$B\to\ds\ellm\nulb$. We perform a fit with these modified
PDFs and take the difference from the nominal fit as a
systematic uncertainty.

\subsubsection{\piz Efficiency and Crossfeed Constraints}
While the systematic uncertainties due to detector
efficiencies (described in more detail in Section~\ref{sub:deteff})
are primarily multiplicative, the efficiencies for \piz
reconstruction have a large impact on the
feed-down efficiencies and therefore the fit yields.
This effect can be enhanced by the fact that the feed-down
constraints are defined as the ratio of the number of events
reconstructed in the $D\ellm$ channel to that in the $\Dstar\ellm$ channel,
which move in opposite directions as the \piz efficiency
is varied.

We generate an ensemble of fits by varying the \piz
efficiency within its uncertainty,
$3.0\%$ per \piz. The resulting
changes in the feed-down constraints for both signal
and background modes are propagated through the signal
fit to estimate the resulting systematic uncertainties.

\subsubsection{$D^{**}\ellm\nulb$ Feed-down}
We assign an additional systematic uncertainty on
$D^{**}\ellm\nulb$ feed-down rates due to the fact
that the \piz mesons involved in feed-down
processes typically have low momentum, while the
$3.0\%$ systematic uncertainty mentioned above is
derived from a control sample with a broad spectrum.
Since we float the constraints describing
$\Dstar\Rightarrow D$ feed-down in the fit, \Dstar
feed-down processes are insensitive
to systematic effects due to the \piz efficiency at low
momentum. The $D^{**}$ feed-down constraints, however, are taken
from simulation and can therefore be affected.

We compare the fitted values of the $\Dstar\Rightarrow D$ feed-down
rates to the simulation to estimate that the efficiency
for low-momentum \piz mesons is correctly modeled to within
$10\%$. We generate an ensemble of fits in which we vary
the \piz reconstruction efficiency $\pm 10\%$ for \piz mesons
with momentum less than $300\ \mevc$. We generate new PDFs
and feed-down constraints which we propagate through the
signal fit to estimate the systematic uncertainties.

\subsubsection{$B\to D^{**}\taum\nutb$ Abundance}
We vary the fraction of $B\to D^{**}\taum\nutb$ events
in the $D^{**}$ samples by generating random numbers
from a Gaussian distribution with mean $1.0$ and
width $0.3$, equivalent to a $\pm 30\%$ variation.
For each test, we generate new PDFs and crossfeed
constraints to estimate the systematic uncertainty.

\subsection{Multiplicative Systematic Uncertainties}
\subsubsection{Monte Carlo Statistics}
The dominant multiplicative systematic uncertainty
is due to limited Monte Carlo statistics. The various
MC samples are independent of one another, so that there is
no cancellation between the signal and normalization.

\subsubsection{Bremsstrahlung and Final-State Radiation}
Based on a control sample of identified electrons and studies
in MC samples, we estimate the uncertainty on reconstruction
efficiency due to Bremsstrahlung and final-state radiation
effects to be $2.1\%$. This uncertainty applies to both
signal and normalization modes, however, and so the effect
on the relative efficiency is expected to cancel. The fractions
of events in which a photon is radiated are nearly the same
between signal and normalization modes, within statistical
precision of $10\%$; we therefore treat the uncertainty
between signal and normalization modes as $90\%$ correlated
to calculate the final systematic uncertainty.

\subsubsection{Detector Efficiencies}\label{sub:deteff}
We estimate systematic uncertainties related to the
detector efficiencies---track and neutral reconstruction
and charged particle identification---by studying these efficiencies
in several control samples in both data and simulation. We correct
the MC efficiencies to match those seen in the data,
and we take the statistical precision of these studies
as an estimate of the systematic uncertainty on absolute
efficiencies.

Since we normalize our signals to
$B\to\ds\ellm\nulb$, we calculate systematic uncertainties
on the relative efficiency, treating uncertainties on
the signal and normalization modes as correlated. The
degree of correlation, and therefore, the degree to which
the uncertainty cancels, is determined by the kinematics
of the two samples. For most of the final state
particles, the kinematic distributions are very similar
between signal and normalization modes and so the
systematic uncertainty cancels almost entirely. For
the charged leptons, however, the momentum spectra are
very different between signal and normalization (see
Fig.~\ref{fig:cln_el}), and so the associated systematic
uncertainty is larger.

\subsubsection{Hadronic Daughter Branching Fractions}
We reconstruct both signal and normalization modes using
the same set of final states, so uncertainties due to the
branching fractions of these states very nearly cancel.
(The \ds momentum spectra are slightly different
between signal and normalization modes, so this cancellation
is not perfect.) We take the uncertainty on each of the
reconstructed \Dstar, $D$, \KS, and \piz
decay modes from~\cite{PDG} and propagate each of these
uncertainties through to the relative efficiency, using
the relative abundance of each decay chain in the
signal and normalization MC samples to determine the
correlation and the degree of cancellation.

\subsubsection{Leptonic $\tau$ Branching Fraction}
The $\tau$ branching fraction $\cal B(\taum\to\ellm\nulb\nut)$
appears only in the denominator of Eq.~\ref{eq:norm} and
therefore contributes a $0.2\%$ systematic uncertainty
on all modes~\cite{PDG} without cancellation.

\section{Results}\label{results}
Table~\ref{tab:results} summarizes the results from two fits, one in
which all four signal yields can vary independently, and the second
\Bm--\Bzb constrained fit with
$R(\Dp)=R(\Dz)$ and $R(\Dstarp)=R(\Dstarz)$. We observe approximately
$67\ B\to D\taum\nutb$ and $101\ B\to\Dstar\taum\nutb$
signal events in this \Bm--\Bzb-constrained fit, corresponding
to signal branching-fraction ratios of $R(D)=(41.6\pm 11.7\pm 5.2)\%$ and
$R(\Dstar)=(29.7\pm 5.6\pm 1.8)\%$, where the first error is
statistical and the second systematic. Normalizing these to
known \Bzb branching fractions,\footnote{We use~\cite{PDG} to
normalize the four individual branching fractions. For
the \Bm--\Bzb-constrained measurement, we use our own
averages of the values in~\cite{PDG}:
${\cal B}(\Bzb\to\Dp\ellm\nulb)=(2.07\pm 0.14)\%$ and
${\cal B}(\Bzb\to\Dstarp\ellm\nulb)=(5.46\pm 0.18)\%$.} we obtain
${\cal B}(B\to D\taum\nutb)=(0.86\pm 0.24\pm 0.11\pm 0.06)\%$ and
${\cal B}(B\to\Dstar\taum\nutb)=(1.62\pm 0.31\pm 0.10\pm 0.05)\%$,
where the third error is from that on the normalization branching
fraction.

Table~\ref{tab:results} also gives the significances of the signal yields. 
The statistical significance is determined from $\sqrt{2\Delta(\ln\cal L)}$,
where $\Delta(\ln\cal L)$ is the change in log-likelihood between the nominal fit and
the no-signal hypothesis. The total significances are determined by
including the systematic uncertainties on the fit yields in quadrature with the statistical errors.
In the \Bm--\Bzb-constrained fit, the signal significances are $3.6\sigma$ and
$6.2\sigma$ for $R(D)$ and $R(\Dstar)$, respectively.

The statistical correlation between $R(D)$ and $R(\Dstar)$ is $-0.51$ in
the \Bm--\Bzb-constrained fit. This correlation is due to the fact
that most of the events at large \mmiss are either $B\to D\taum\nutb$
or $B\to\Dstar\taum\nutb$ signal events, and increasing either of
the two signal yields in the fit necessarily decreases the other.
The systematic uncertanties have a correlation of $-0.03$ between $R(D)$ and $R(\Dstar)$; most
of the systematic uncertainties have large negative correlations
for the same reason that the statistical uncertainty does, but
the combinatorial background uncertainty affects both signal
yields in a coherent manner and so contributes a large positive
correlation. The sum of the two branching fractions,
taking all correlations into account, is
${\cal B}(B\to\ds\taum\nutb)=(2.48\pm 0.28\pm 0.15\pm 0.08)\%$.

\begin{table*}
\caption{Results from fits to data:
the signal yield ($N_\mathrm{sig}$), the yield of normalization $B\to\ds\ellm\nulb$ events ($N_\mathrm{norm}$),
the relative systematic error due to the fit yields [$(\Delta R/R)_\mathrm{fit}$], the
relative systematic error due to the efficiency ratios [$(\Delta R/R)_\varepsilon$],
the branching-fraction ratio ($R$), the absolute
branching fraction ($\cal B$), and the total and statistical signal significances
($\sigma_\mathrm{tot}$ and $\sigma_\mathrm{stat}$).
The first two errors on $R$
and $\cal B$ are statistical and systematic, respectively; the third error on $\cal B$
represents the uncertainty on the normalization mode.
The last two rows show the results of the fit with the \Bm--\Bzb constraint applied,
where $\cal B$ is expressed for the \Bzb. The statistical correlation between
$R(D)$ and $R(\Dstar)$ in this fit is $-0.51$.
}
\label{tab:results}
\begin{tabular}{l l r@{$\pm$}l r@{$\pm$}l c c r@{$\pm$}r@{$\pm$}r l@{$\pm$}l@{$\pm$}l@{$\pm$}l l}\\ \hline\hline
\multicolumn{2}{l}{Mode} & \multicolumn{2}{c}{$N_\mathrm{sig}$} &
  \multicolumn{2}{c}{$N_\mathrm{norm}$} &
  \multicolumn{1}{c}{$(\Delta R/R)_\mathrm{fit}$} &
  \multicolumn{1}{c}{$(\Delta R/R)_\varepsilon$} &
  \multicolumn{3}{c}{$R$} &
  \multicolumn{4}{c}{$\cal B$} & \multicolumn{1}{c}{$\sigma_\mathrm{tot}$} \\
\multicolumn{6}{c}{} &
  \multicolumn{1}{c}{[\%]} &
  \multicolumn{1}{c}{[\%]} &
  \multicolumn{3}{c}{[\%]} & \multicolumn{4}{c}{[\%]} &
  \multicolumn{1}{c}{$(\sigma_\mathrm{stat})$} \\ \hline
\Bm & $\to\Dz\taum\nutb$ & $35.6$ & $ 19.4$ & $347.9$ & $23.1$ & $15.5$ & $1.6$ &
  $31.4$ & $17.0$ & $4.9$ & $0.67$ & $0.37$ & $0.11$ & $0.07$ & $1.8\ (1.8)$ \\
\Bm & $\to\Dstarz\taum\nutb$ & $92.2$ & $19.6$ & $1629.9$ & $63.6$ & $9.7$ & $1.5$ &
  $34.6$ & $7.3$ & $3.4$ & $2.25$ & $0.48$ & $0.22$ & $0.17$ & $5.3\ (5.8)$ \\
\Bzb & $\to\Dp\taum\nutb$ & $23.3$ & $7.8$ & $150.2$ & $13.3$ & $13.9$ & $1.8$ &
  $48.9$ & $16.5$ & $6.9$ & $1.04$ & $0.35$ & $0.15$ & $0.10$ & $3.3\ (3.6)$ \\
\Bzb & $\to\Dstarp\taum\nutb$ & $15.5$ & $7.2$ & $482.3$ & $25.5$ & $3.6$ & $1.4$ &
  $20.7$ & $9.5$ & $0.8$ & $1.11$ & $0.51$ & $0.04$ & $0.04$ & $2.7\ (2.7)$ \\ \hline
$B$ & $\to D\taum\nutb$ & $66.9$ & $18.9$ & $497.8$ & $26.4$ & $12.4$ & $1.4$ &
  $41.6$ & $11.7$ & $5.2$ & $0.86$ & $0.24$ & $0.11$ & $0.06$ & $3.6\ (4.0)$ \\
$B$ & $\to\Dstar\taum\nutb$ & $101.4$ & $19.1$ & $2111.5$ & $ 68.1$ & $5.8$ & $1.3$ &
  $29.7$ & $5.6$ & $1.8$ & $1.62$ & $0.31$ & $0.10$ & $0.05$ & $6.2\ (6.5)$ \\ \hline\hline
\end{tabular}
\end{table*}

\begin{figure}
\includegraphics[width=3.1in]{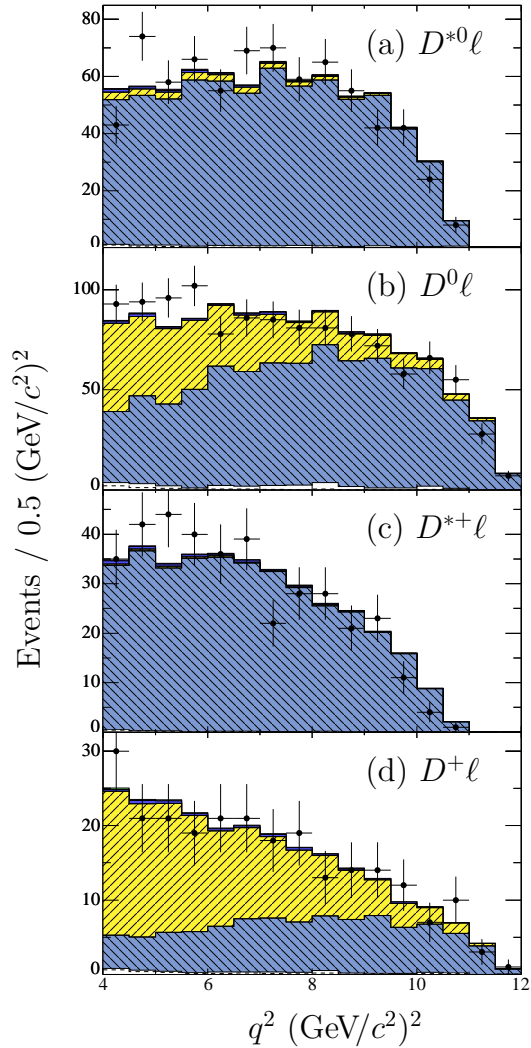}
\caption{(Color online) $q^2$ distributions of events in the
four final states
$\Dstarz\ell^-$, $\Dz\ell^-$, $\Dstarp\ell^-$, and $\Dp\ell^-$,
shown in the normalization region, $\mmiss<1\ (\gevccnosp)^2$.
The data are shown as points with error bars. The shaded
histograms are taken from MC samples with normalizations
from the fit to data.
The components are shaded as in Fig.~\ref{fig:fit}.}
\label{fig:q2norm}
\end{figure}

\begin{figure}
\includegraphics[width=3.1in]{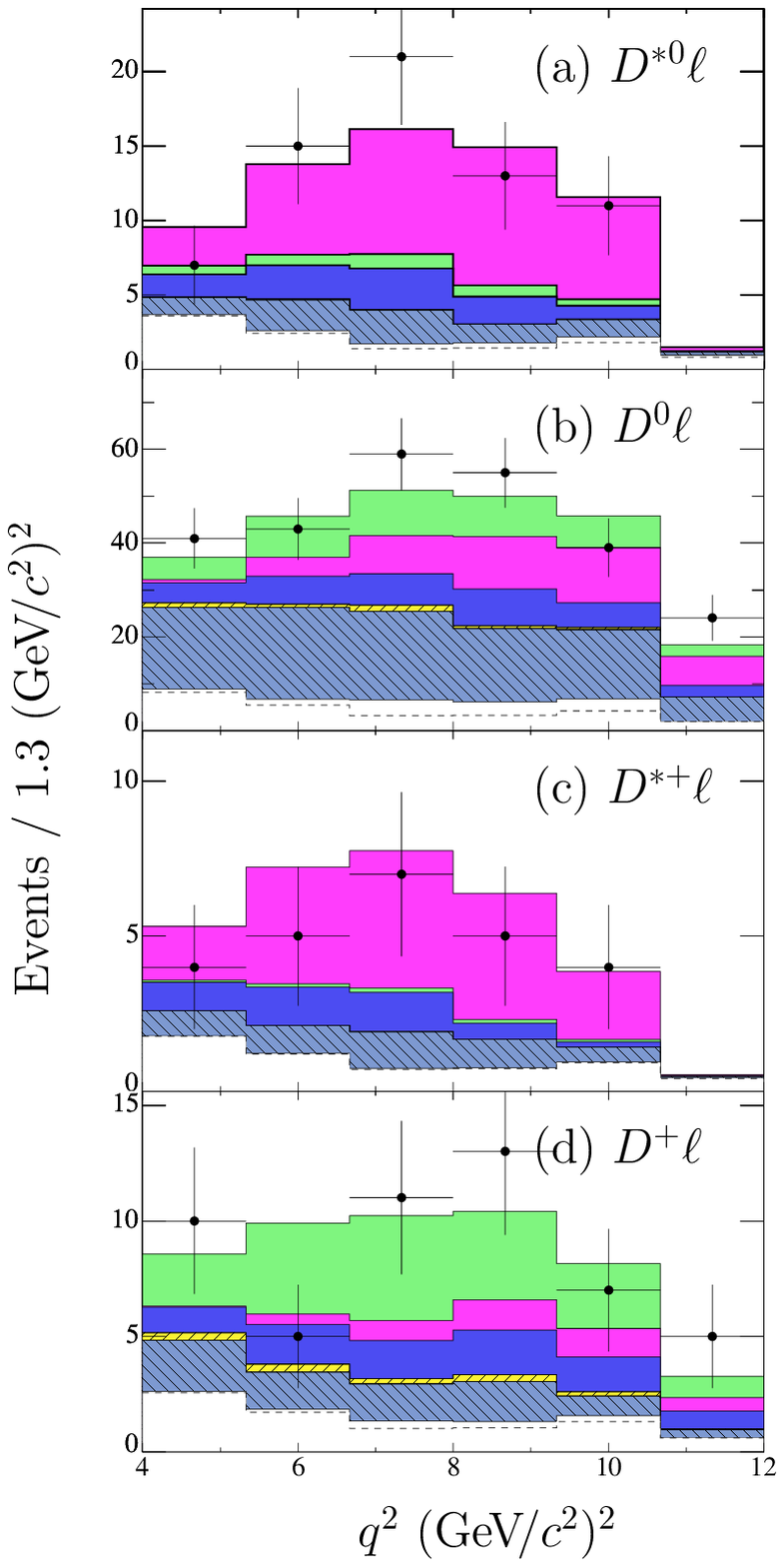}
\caption{(Color online) $q^2$ distributions of events in the
four final states
$\Dstarz\ell^-$, $\Dz\ell^-$, $\Dstarp\ell^-$, and $\Dp\ell^-$,
shown in the signal region, $\mmiss>1\ (\gevccnosp)^2$.
The data are shown as points with error bars. The shaded
histograms are taken from MC samples with normalizations
from the fit to data.
The components are shaded as in Fig.~\ref{fig:fit}.}
\label{fig:q2sig}
\end{figure}

Figures~\ref{fig:q2norm} and \ref{fig:q2sig} show the observed
$q^2$ distributions in the four signal channels in the low and
high \mmiss regions, respectively. The histograms in these
figures are taken from MC samples of the various components,
with each component scaled to match the yield in the
\Bm--\Bzb-constrained fit; since $q^2$ is not a fit
variable, we cannot show a projection of a continuous
PDF as was done in Figs.~\ref{fig:fit}--\ref{fig:pstardss}.
As before, we observe good agreement between the data
and the expectation from simulation, in both the low and
high \mmiss regions. Since the $q^2$ distribution is
highly dependent on the form factor model, we note that
the CLN model describes both normalization and signal
events within the available statistics.

Table~\ref{tab:checks} summarizes the results of several crosschecks,
including splitting up the sample 
according to lepton flavor, lepton charge, and data-taking period.
We have done these checks by performing ``cut-and-count'' analyses,
both in the data and in simulated event samples. In all cases, the
results in data are consistent with our expectations from simulation.
The first row in this table shows the fraction of events
with muon candidates in data and simulation, both for the full event
sample and for the signal-sensitive region in \mmiss. Electron
identification is more efficient than muon ID, which is why the
muon fraction in the final sample is less than 50\%, and, at lower
momenta (which generally correspond to larger \mmiss), this efficiency difference
is more pronounced; in both cases, however, the muon abundance is
well-modelled by the simulation. The next row shows the fraction of
positively-charged lepton candidates (versus negatively-charged
candidates), and all samples are consistent with the expected
50/50 split. The last row shows the fraction of events recorded
during the Run~4 \babar\ data-taking period; Run~4 had significantly
different accelerator background conditions from Runs~1--3, which
could affect missing-energy analyses.
The fraction of events in the Run~4 subsample is
consistent with expectations: Run~4 makes up 47\%
of the total luminosity.

\begin{table}
\caption{Crosscheck studies, splitting the data according to
lepton flavor, lepton charge, and running period.
The first row shows the fraction of events with muon candidates
for both data and MC samples, for both the full event sample
and for the signal-sensitive region $\mmiss > 1\ (\gevccnosp)^2$.
The second row shows fractions of events with positively charged
lepton candidates, and the third row shows the fractions of
events recorded in Run 4. In all cases, the data are consistent
with the simulation and with expectations.}
\label{tab:checks}
\begin{tabular}{l l l l l l}\\ \hline\hline
 & \multicolumn{2}{c}{Full sample} & & \multicolumn{2}{c}{High \mmiss}\\ \cline{2-3}\cline{5-6}
Sample & \multicolumn{1}{c}{$f_\mathrm{data}$ (\%)} & \multicolumn{1}{c}{$f_\mathrm{MC}$ (\%)} & & \multicolumn{1}{c}{$f_\mathrm{data}$ (\%)} & \multicolumn{1}{c}{$f_\mathrm{MC}$ (\%)}\\ \hline
$\mu$   & $40.0\pm 0.9$ & $40.9\pm 0.4$ & & $30.7\pm 2.3$ & $31.9\pm 1.0$\\
$\ellp$ & $50.2\pm 1.0$ & $49.2\pm 0.4$ & & $49.3\pm 2.5$ & $48.9\pm 1.0$\\
Run 4   & $44.6\pm 1.0$ & $47.6\pm 0.5$ & & $46.8\pm 2.5$ & $48.5\pm 1.2$\\ \hline\hline
\end{tabular}
\end{table}

We estimate the goodness of fit using an ensemble of simulated
experiments. We generate 1000 event samples, using the
nominal PDFs for the fit to data and event yields based on the
\Bm--\Bzb-constrained fit to data. We fit each of these samples
both with and without the \Bm--\Bzb constraints and study the distribution
of \nll in these fits.

\begin{figure}
\includegraphics[width=3.1in]{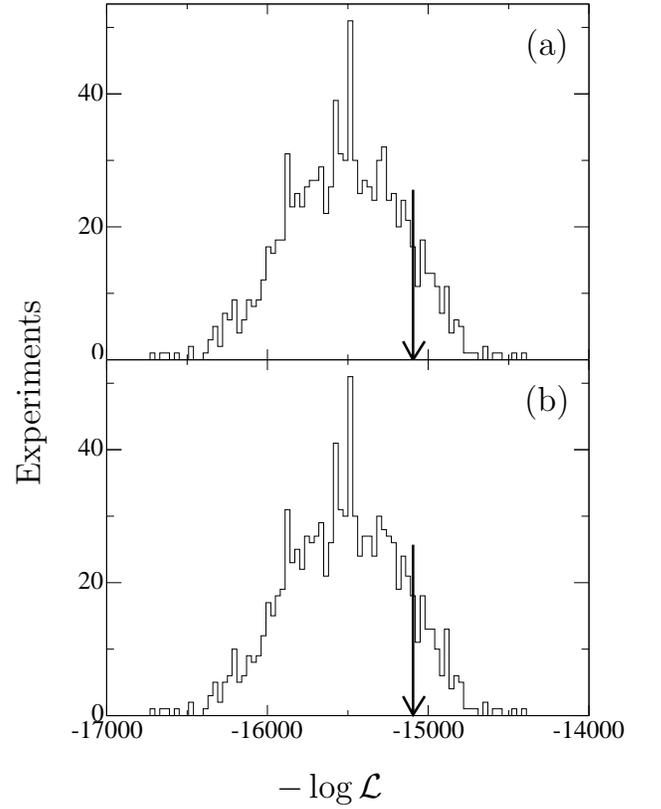}
\caption{Distribution of \nll from simulated experiments, showing (a) the unconstrained fit and
(b) the \Bm--\Bzb constrained fit. The observed values of \nll in the
fit to data are indicated with arrows. The fraction of experiments
with \nll larger than the observed value is used to estimate the goodness of fit.
}
\label{fig:nll}
\end{figure}

Figure~\ref{fig:nll} shows the distribution of \nll for the
two ensembles of fits. In both
cases, the value of \nll obtained in the fit to data is
indicated with an arrow, and, in both cases, this value is
found within the central part of the Monte Carlo distribution,
indicating a good fit.
In the unconstrained fit, $11.7\%$ of the simulated experiments have a value
of \nll greater than the value observed in data, corresponding to the
probability that we expect to observe a fit as bad, or worse, than the
one actually observed. This probability is large, indicating
an acceptable goodness of fit. The corresponding probability for the
\Bm--\Bzb constrained fit is $11.8\%$, also large.

\section{Conclusions}
We have presented measurements of the branching fractions for the decays
$B\to D\taum\nutb$ and $B\to\Dstar\taum\nutb$, determined relative to the
corresponding decays to light leptons. We measure the branching-fraction
ratios for four individual $\ds$ states, as well as two 
\Bm--\Bzb-constrained ratios

\begin{eqnarray}
R(\Dz)    &=& (31.4\pm 17.0\pm 4.9)\% \nonumber\\
R(\Dstarz)&=& (34.6\pm 7.3\pm 3.4)\% \nonumber\\
R(\Dp)    &=& (48.9\pm 16.5\pm 6.9)\% \nonumber\\
R(\Dstarp)&=& (20.7\pm 9.5\pm 0.8)\% \nonumber\\
R(D)      &=& (41.6\pm 11.7\pm 5.2)\% \nonumber\\
R(\Dstar) &=& (29.7\pm 5.6\pm 1.8)\% \nonumber~,
\end{eqnarray}

\noindent where the first uncertainty is statistical and the second is systematic.
The significances of these signals are $1.8\sigma$, $5.3\sigma$,
$3.3\sigma$, $2.7\sigma$, $3.6\sigma$, and $6.2\sigma$, respectively.
The statistical and systematic uncertainties on $R(D)$ and $R(\Dstar)$
have correlations of $-0.51$ and $-0.03$, respectively.

From these branching-fraction ratios and known branching fractions
of the normalization modes $B\to\ds\ellm\nulb$, we derive
the absolute branching fractions

\begin{eqnarray}
{\cal B}(\Bm\to\Dz\taum\nutb)    &=&(0.67\pm 0.37\pm 0.11\pm 0.07)\% \nonumber\\
{\cal B}(\Bm\to\Dstarz\taum\nutb)&=&(2.25\pm 0.48\pm 0.22\pm 0.17)\% \nonumber\\
{\cal B}(\Bzb\to\Dp\taum\nutb)    &=&(1.04\pm 0.35\pm 0.15\pm 0.10)\% \nonumber\\
{\cal B}(\Bzb\to\Dstarp\taum\nutb)&=&(1.11\pm 0.51\pm 0.04\pm 0.04)\% \nonumber\\
{\cal B}(B\to D\taum\nutb)     &=&(0.86\pm 0.24\pm 0.11\pm 0.06)\% \nonumber\\
{\cal B}(B\to\Dstar\taum\nutb) &=&(1.62\pm 0.31\pm 0.10\pm 0.05)\% \nonumber~,
\end{eqnarray}
\noindent where the third uncertainty reflects that of the normalization
mode branching fraction.

The measurement of ${\cal B}(\Bzb\to\Dstarp\taum\nutb)$ 
is consistent with the Belle result~\cite{Belle}.
The branching-fraction ratios $R(D)$ and $R(\Dstar)$
are about $1\sigma$ higher than the SM predictions but, given the
uncertainties, there is still room for a sizeable non-SM contribution. 

We have also presented distributions of the lepton momentum \pstarl
and the squared momentum transfer $q^2$ for $B\to\ds\taum\nutb$
events. In all cases, these distributions are consistent with
expectations based on the SM and the CLN form factor model with measured form factors.

We are grateful for the 
extraordinary contributions of our \pep2\ colleagues in
achieving the excellent luminosity and machine conditions
that have made this work possible.
The success of this project also relies critically on the 
expertise and dedication of the computing organizations that 
support \babar.
The collaborating institutions wish to thank 
SLAC for its support and the kind hospitality extended to them. 
This work is supported by the
US Department of Energy
and National Science Foundation, the
Natural Sciences and Engineering Research Council (Canada),
the Commissariat \`a l'Energie Atomique and
Institut National de Physique Nucl\'eaire et de Physique des Particules
(France), the
Bundesministerium f\"ur Bildung und Forschung and
Deutsche Forschungsgemeinschaft
(Germany), the
Istituto Nazionale di Fisica Nucleare (Italy),
the Foundation for Fundamental Research on Matter (The Netherlands),
the Research Council of Norway, the
Ministry of Education and Science of the Russian Federation, 
Ministerio de Educaci\'on y Ciencia (Spain), and the
Science and Technology Facilities Council (United Kingdom).
Individuals have received support from 
the Marie-Curie IEF program (European Union) and
the A. P. Sloan Foundation.

\end{document}